\documentclass[%
 reprint,
superscriptaddress,
 amsmath,amssymb,
 aps,
]{revtex4-1}

\usepackage{graphicx}
\usepackage{dcolumn}
\usepackage{bm}
\usepackage[mathlines]{lineno}

\usepackage{pifont}

\def\braf{(\kern-3pt(}     \def\ketf{)\kern-3pt)}

\begin{document}

\preprint{APS/123-QED}

\title{Quantum Plasmonics with multi-emitters: Application to adiabatic control}

\author{A. Castellini}
\address{\small Dipartimento di Fisica dell'Universit\`a di Palermo, Via Archirafi 36, I-90123 Palermo, Italy}
\address{\small Laboratoire Interdisciplinaire Carnot de Bourgogne, CNRS UMR 6303, Universit\'e Bourgogne Franche-Comt\'e BP 47870, 21078 Dijon, France
}
\address{\small INFN Sezione di Catania, Catania, Italy}
\author{H. R. Jauslin}%
\address{\small Laboratoire Interdisciplinaire Carnot de Bourgogne, CNRS UMR 6303, Universit\'e Bourgogne Franche-Comt\'e BP 47870, 21078 Dijon, France
}
\author{B. Rousseaux}%
\address{\small Laboratoire Interdisciplinaire Carnot de Bourgogne, CNRS UMR 6303, Universit\'e Bourgogne Franche-Comt\'e BP 47870, 21078 Dijon, France
}
\author{D. Dzsotjan}%
\address{\small Wigner Research Center for Physics, Hungarian Academy of Sciences, Konkoly-Thege Miklos ut 29-33, H-1121 Budapest, Hungary
}
\author{G. Colas des Francs}%

\address{\small Laboratoire Interdisciplinaire Carnot de Bourgogne, CNRS UMR 6303, Universit\'e Bourgogne Franche-Comt\'e BP 47870, 21078 Dijon, France
}


\author{A. Messina}
\address{\small Dipartimento di Fisica dell'Universit\`a di Palermo, Via Archirafi 36, I-90123 Palermo, Italy}
\address{\small INFN Sezione di Catania, Catania, Italy}

\author{S. Gu\'erin}
\email{sguerin@u-bourgogne.fr}
\address{\small Laboratoire Interdisciplinaire Carnot de Bourgogne, CNRS UMR 6303, Universit\'e Bourgogne Franche-Comt\'e BP 47870, 21078 Dijon, France
}


\date{\today}

\begin{abstract}
We construct mode-selective effective models describing the interaction of $N$ quantum emitters (QEs) with the localised surface plasmon polaritons (LSPs) supported by a spherical metal nanoparticle (MNP) in an arbitrary geometric arrangement of the QEs. We develop a general formulation in which the field response in the presence of the nanosystem can be decomposed into orthogonal modes with the spherical symmetry as an example.
We apply the model in the context of quantum information, investigating on the possibility of using the LSPs as mediators of an efficient control of population transfer between two QEs. We show that a Stimulated Raman Adiabatic Passage configuration allows such a transfer via a decoherence-free dark state when the QEs are located on the same side of the MNP and very closed to it, whereas the transfer is blocked when the emitters are positioned at the opposite sides of the MNP.
We explain this blockade by the destructive superposition of all the interacting plasmonic modes. 
\end{abstract}

\maketitle
\section{INTRODUCTION} \label{Intr}
Plasmonics offers new scenarios for classical and quantum manipulations of light at the nanoscale \cite{Leru,Maierbook}. It can be implemented via metallic nano-particles (MNPs) supporting propagating and localised surface plasmon polaritons (SPPs and LSPs, respectively) at dielectric/metallic interfaces \citep{Zayats,Ozbay189}. Plasmonic quantum electrodynamics (PQED) \cite{PQED,CdF} is the natural generalization of the principles of quantum optics for cavity quantum electrodynamics (cQED) \cite{CQED}, enabling to treat the coupling between quantum emitters (QEs) and the modes of the SPP field localised in subdiffraction volumes. A model has been initially proposed \cite{Knoll,David1,uno,Zubairy}, where all the electromagnetic excitations (corresponding to LSPs) of the system are described by collective field operators, exploiting the theory of quantization of the electromagnetic field in dispersive and absorbing media \cite{Suttorp-vanWonderen-2004-EPL,Knoll,HuttandBarn,Philbin}. 
When the QEs are placed very close to the MNP's surface (typically less than 10 nm), the electromagnetic local density of states (LDOS) of the LSP are magnified and can lead to a strong coupling \cite{Zubairy,David}. The role of the lower and higher order LSP modes in the interaction with QEs depends on their positions around the metal \cite{Zubairy}. 
The quasi-degenerate higher order modes can collectively act, in the interaction with QEs, as a pseudomode rather than a Markovian bath \cite{Delga1}. As a result, effective few-states Hamiltonians have been phenomenologically proposed to describe the weak and strong coupling regimes of layer or ring configurations of $N$ QEs with the dipolar mode and with this pseudomode \cite{Delga1,Delga}. In order to capture the electromagnetic degrees of freedom playing the major role in the coupled dynamics (bright LSP contributions), cQED-like hamiltonian models have been derived \citep{David}. Mode-selective LSP field operators, i.e. associated to the different plasmonic resonances, substitute thus the collective electromagnetic field operators to describe the interaction of the different LSP modes with QEs \cite{David}. The deeper understanding of these matter-plasmonic platforms and their nanometric dimensions stimulated the interest for their use as devices for classical and quantum information processes \cite{Lukin,Susa}. 

However, the plasmonic quality factors are much smaller than the cavity's ones, which prevents in principle SPPs from carrying some information over longer distances.
%
In \citep{Benjamin}, it has been shown that a free-decoherence quantum channel linking two not-directly interacting QEs can be created exploiting the mediation of LSPs of a spherical MNP, in a specific geometrical configuration. In this case a Stimulated Raman adiabatic Passage (STIRAP) \citep{Vit15,Vitanov} allows a population transfer between two states of the total system, resulting in a population exchange or in an entanglement between the two QEs, however for emitters aligned at the same side of the sphere \citep{Benjamin}.

In this paper, we develop a complete procedure to obtain a cQED-like effective model of a system of $N$ QE's coupled to LSPs of a spherical MNP in an arbitrary geometric arrangement. It is formulated in a more general way, when the Green tensor associated to the field response in presence of the nanosystem can be decomposed into orthogonal modes with the spherical symmetry as an example.
The mathematical technique used in this paper completes and improves preceding methods \citep{David} in generalizing and applying the L\"{o}wdin orthonormalization \cite{Lowdin} of a set of mode-selective bright LSP field operators. This gives us compact analytical expressions for the QE / LSP coupling constants, for any possible angular disposition of a given number of emitters around the sphere.

We next exploit the derived model numerically by simulating a STIRAP process in the case of two QEs targeting an adiabatic population transfer between two states of the global system, as an exchange between their population. We analyse the dependence of the transfer efficiency 
with respect to (i) the angular distance between the emitters and (ii) the distance from the metallic surface via the number of effectively involved plasmonic modes.
The existence of a blockade of the population transfer for two QEs located at the opposite sides of the MNP is explained by a destructive interference of the plasmonic modes. The population transfer is achieved when he QEs are located on the same side of the MNP.

This article is organized as follows: in Sec. \ref{II} we construct the continuous and discrete effective mode-selective hamiltonian models describing the interaction of $N$ QEs with a spherical MNP, for any possible overlap of the plasmonic modes excited by different emitters. In Sec \ref{III} we exploit the derived discrete effective model to show under which geometrical and parametric conditions a STIRAP process can be numerically implemented. Finally, In Sec. \ref{IV} we summarize the obtained results and the perspectives.

\section{EFFECTIVE models for the QE / LSP system}\label{II}

\begin{center}
\begin{figure}
\includegraphics[scale=0.4]{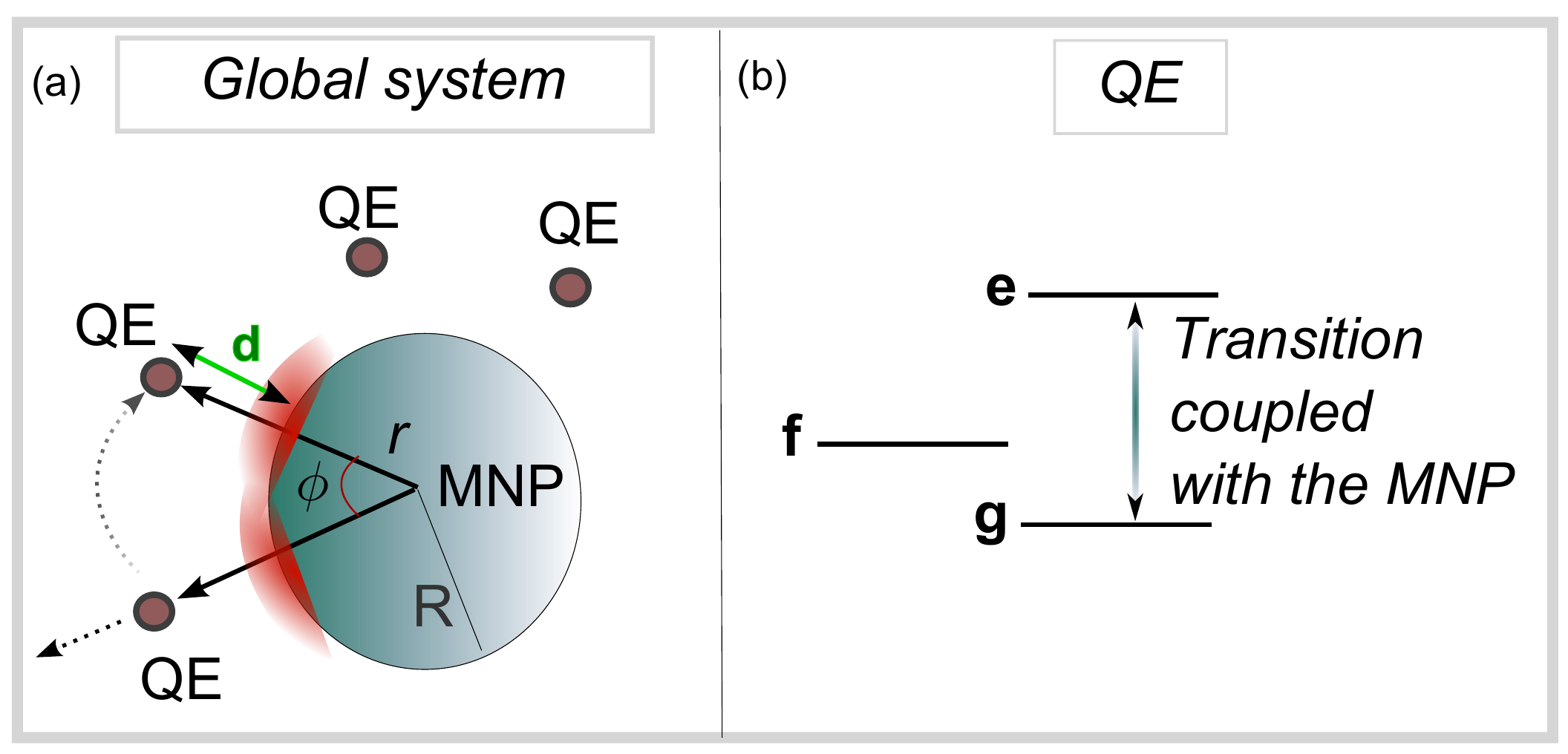}
\caption{(a): Global system composed by a set of QEs placed at a distance $d$ ($r$) from the metallic surface (center) of a MNP of radius $R$. The interaction of each QE with the MNP excites different LSPP modes (represented by the red clouds on the MNP's surface). The angular distance between any two QEs is indicated by $\phi$. (b) Scheme of the 3-level $\Lambda$ structure of each QE, composed by two metastable states, $|g\rangle$ and $|f\rangle$, and an excited state $|e\rangle$. The transition $|e\rangle - |g\rangle$ is coupled with the MNP.}
\label{fig:System}
\end{figure}
\end{center}
We want to describe the dynamics of a system composed by (see Fig. \ref{fig:System})
\begin{itemize}
\item a hybrid \textit{matter-electromagnetic field} system, which originates from the interaction of the free electromagnetic field with the normal modes associated to the collective surface oscillations of the conduction electrons in a spherical MNP. The latter is characterized by a radius $R$ and by a dielectric permittivity $\epsilon_m(\omega)$. 
The excitation associated to the global system is annihilated by the bosonic field operator $\hat{\mathbf{f}}_{\omega}(\mathbf{r})$  \cite{Knoll}. 
\item $N$ identical 3-level QEs, each of which is composed by two metastable states, $|g\rangle$ and $|f\rangle$, and an excited one $|e\rangle$. Each QE is in $\Lambda$ configuration, that is $E_g\le E_f<E_e$, and is placed at a position $\mathbf{r}_i$ from the center of the MNP, corresponding to a distance $d_i$, $i=1,..,N$, from its surface.
\end{itemize}
The distance between the QEs is such that their direct coulombic interaction is negligible compared to that mediated by the plasmonic field \cite{Zubairy,David,Delga}. Each QE is coupled to the MNP through the transition $|e\rangle - |g\rangle$. The transition $|f\rangle - |g\rangle$ is forbidden and the transition $|e\rangle -|f\rangle$ will be used only via a control field.

The results of the paper can be summarized as follows:
\begin{enumerate}
\item We construct a mode-selective effective Hamiltonian model describing the global dynamics, which involves only the relevant degrees of freedom. We derive compact analytical expressions of the QE / LSP coupling constants for any geometrical arrangement of the QEs around the sphere, improving previous approaches \citep{Delga1,David};

\item We study the efficiency of the population transfer in a STIRAP configuration, in the simplest case of $N=2$, between two metastable states of the global system, which would result in an exchange of population between the two QEs. We analyse the dependence of the process on the angular distance of the two QEs, on the number of involved LSP modes and on the distance of the QEs from the metallic surface, which completes the analysis presented in \cite{Benjamin}.
\end{enumerate}


\noindent We first omit the metastable state $|f\rangle$, which will be involved in the control process in Section III.
\noindent We start from a rotating wave approximation (RWA) model \cite{Knoll,David1,uno,Zubairy,David}:
\begin{equation}\begin{split}
\hat{H}=&\int_0^{+\infty} d\omega \ \hbar \omega \int d^3r \ \hat{\mathbf{f}}^{\dagger}_{\omega}(\mathbf{r})\cdot\hat{\mathbf{f}}_{\omega}(\mathbf{r})+\sum_{l=1}^N \hbar \omega_{eg}\hat{\sigma}_{ee}^{(l)}\\
&-\sum_{l=1}^N \hat{\sigma}_{eg}^{(l)} \otimes \int_0^{+\infty} d\omega \ \mathbf{d}_{eg}^{(l)} \cdot \hat{\mathbf{E}}^{+}_{\omega}(\mathbf{r}_l)+H.c.,
\end{split}
\label{Hcg}
\end{equation}
with the emitter operators $\hat\sigma_{ij}= \vert i\rangle \langle j\vert$, $\mathbf{d}_{eg}^{(l)}$ the dipole moment at the transition $e-g$ of the $l^{th}$ emitter, and where the components of the elementary excitation operator $\mathbf{\hat{f}}_{\omega}(\mathbf{r})$ obey the following commutation relations:
\begin{equation}
[\hat{f}_{\omega}^{i}(\mathbf{r}),\hat{f}_{\omega'}^{j\, \dagger}(\mathbf{r'})]=\delta(\omega-\omega')\delta(\mathbf{r}-\mathbf{r}')\delta_{ij}.
\label{commf}
\end{equation}
The first term of \eqref{Hcg} represents the free energy of all the plasmonic modes supported by the MNP.
The last one is the interaction energy of the $l^{\text{th}}$ QE, at the position $\mathbf{r}_l$, with the plasmonic electric field $\hat{\mathbf{E}}_{\omega}(\mathbf{r}_l)=\hat{\mathbf{E}}^+_{\omega}(\mathbf{r}_l)+H.c.$.

In the following, we derive the effective model for many emitters using a mode-selective quantization in the Green tensor formalism, similarly as in \citep{David}, but in a more direct way, without expanding the elementary excitation operators $\mathbf{\hat{f}}_{\omega}(\mathbf{r})$ in the spherical vector harmonics basis. Only the Green tensor is expanded in this basis. This gives an Hamiltonian in terms of creation and annihilation operators of a plasmon polariton in each LSP mode (dipolar, quadrupolar,...) excited by the different emitters.
The presence of many emitters leads to overlapping non-orthogonal modes which have to be orthogonalized into bright and dark modes according to the effective interaction between the plasmon polaritons and the QEs. We describe a global orthogonalization procedure based on L\"owdin's methods
\cite{Lowdin-orthogonalization-1950} 
\cite{Lowdin-orthogonalization-1966} 
\cite{Lowdin-book}.

Formally, the dark/bright decomposition can be summarized as follows: we identify and eliminate their contribution from the total Hamiltonian:
\begin{equation}
\label{Htot}
\hat H=\hat{H}_{\text{bright}}^0+\hat{H}_{\text{dark}}^0+\hat{H}_{\text{QEs}}^0+\hat{H}_{\text{bright/QEs}}.
\end{equation}
Since we have $[\hat H,\hat{H}_{\text{dark}}^0]=0$, the dark states are not coupled with the other states and are not affected by the interaction with the emitters, which allows one to drop the $\hat{H}_{\text{dark}}^0$ term in \eqref{Htot}:
\begin{equation} 
\label{Hred}
\hat H=\hat{H}_{\text{bright}}^0+\hat{H}_{\text{QEs}}^0+\hat{H}_{\text{bright/QEs}}.
\end{equation}
\subsection{Mode-selective quantization in the Green tensor formalism}
The electric field in \eqref{Hcg} can be expressed in terms of the inner product of the dyadic Green's function $\mathbf{G}_\omega(\mathbf{r}_l,\mathbf{r})$ with the elementary excitation $\mathbf{\hat{f}}_{\omega}(\mathbf{r})$ \citep{Knoll,David,Delga}.  We consider the case where the Green function can be decomposed as a sum over a discrete index 
\begin{equation}
\bar{\bar{G}}_{\omega}(\mathbf{r},\mathbf{r}')=\sum_n \bar{\bar{G}}_{\omega,n}(\mathbf{r},\mathbf{r}').
\label{G_n_def}
\end{equation} 
as it is the case e.g.  with a multipole expansion in models with spherical symmetry \cite{David,Benjamin}, where the index $n$ is the radial harmonic index. It is referred below to as mode index.

The interaction part of Eq. \eqref{Hcg} can be decomposed into the modes as
\begin{equation}
\hat{H}_{\text{int}}=-i\hbar\sum_{l=1}^N\sum_n\hat\sigma_{eg}^{(l)}\otimes \int d\omega \int d^3r\ \mathbf{g}_{\omega,n}(\mathbf{r}_l,\mathbf{r})\cdot \hat{\mathbf{f}}_{\omega}(\mathbf{r})
\label{Hint1}
\end{equation}
with the vector $\mathbf{g}_{\omega,n}(\mathbf{r}_l,\mathbf{r})$ defined by the inner product of the dipole moment with the dyadic Green's function:
\begin{equation}\begin{split}
\mathbf{g}_{\omega,n}(\mathbf{r}_l,\mathbf{r})&=\sqrt{\dfrac{1}{\hbar \pi \epsilon_0}}\dfrac{\omega^2}{c^2}\sqrt{\epsilon_{\omega}''(\mathbf{r})}\ \mathbf{d}_{eg}^{(l)}\cdot\bar{\bar{G}}_{\omega,n}(\mathbf{r}_l,\mathbf{r})\\
&=\sqrt{\dfrac{1}{\hbar \pi \epsilon_0}}\dfrac{\omega^2}{c^2}\sqrt{\epsilon_{\omega}''(\mathbf{r})}\  \bar{\bar{G}}^T_{\omega,n}(\mathbf{r}_l,\mathbf{r})\cdot\mathbf{d}_{eg}^{(l)},
\end{split}
\label{defg}
\end{equation}
and $\epsilon_{\omega}''(\mathbf{r})$ the imaginary part of the electric permittivity.
For the following development we make the general assumption that the separation into modes of the Green function 
 \eqref{G_n_def} is such that the coefficients \eqref{defg} satisfy the orthogonality  relation
 \begin{equation}
\label{am-orthogonality}
\int d^3r~ {\mathbf{g}}^{*}_{\omega,n}(\mathbf{r}_l,\mathbf{r}) \cdot {\mathbf{g}}_{\omega',n'}(\mathbf{r}_{l'},\mathbf{r})  =0, \qquad {\rm for~} n\neq n'
\end{equation}
 This is the case e.g. for a system with spherical symmetry \citep{David}. 
 The motivation for the separation into modes is that, under suitable conditions, the systems has resonances that produce strong couplings with the emitters at a particular frequency. The resonant frequency  dominates the dynamics for the corresponding mode, and thus the integral over the frequencies $\int d\omega$ appearing in the Hamiltonian can be approximated by a discrete sum, where each term corresponds to a mode $n$. The approximation is completed by the inclusion of an exponential relaxation term for each mode, that takes into account effectively the continuum of frequencies around the resonant one. 

We introduce the mode-selective LSP \textit{bright operators} \citep{David,Benjamin}
\begin{align}
\hat{a}_{\omega,n}(\mathbf{r}_l)
&=\int d^3r \ \mathbf{h}_{\omega,n}(\mathbf{r}_l,\mathbf{r})\cdot \hat{\mathbf{f}}_{\omega}(\mathbf{r})
\label{a_def}
\end{align}
with the vectors $\mathbf{h}_{\omega,n}(\mathbf{r}_l,\mathbf{r}) \in L_2(\mathbb{R}^3,\mathbb{C}^3)$:
\begin{equation}
\mathbf{h}_{\omega,n}(\mathbf{r}_l,\mathbf{r})=\mathbf{g}_{\omega,n}(\mathbf{r}_l,\mathbf{r})/\kappa_{\omega,n}(\mathbf{r}_l),
\label{H_def}
\end{equation}
and
\begin{subequations}
\begin{align}
\label{Kappa_def}
|\kappa_{\omega,n}(\mathbf{r}_l)|^2&=\int d^3r \ \mathbf{g}_{\omega,n}(\mathbf{r}_l,\mathbf{r})\cdot \mathbf{g}^{\ast}_{\omega,n}(\mathbf{r}_l,\mathbf{r}),\\
&=\dfrac{1}{\hbar \pi \epsilon_0}\dfrac{\omega^2}{c^2}\mathbf{d}_{eg}\cdot \text{Im} \left[\bar{\bar G}_{\omega,n}(\mathbf{r}_l,\mathbf{r}_l)\right]\cdot\mathbf{d}_{eg}^{\ast},
\label{Kappa}
\end{align}
\end{subequations}
satisfying the normalization
\begin{align}
\int d^3r \ \mathbf{h}_{\omega,n}(\mathbf{r}_l,\mathbf{r})\cdot \mathbf{h}^{\ast}_{\omega,n'}(\mathbf{r}_l,\mathbf{r})&:=&\braf \mathbf{h}_{\omega,n'}(\mathbf{r}_l),\mathbf{h}_{\omega,n}(\mathbf{r}_l)\ketf\nonumber\\
=\delta_{nn'},
\end{align}
via the identity for non-magnetic material
\cite{Knoll} \cite{Dung-Knoll-Welsch_PRA-1998}
\begin{align}
\int {\mathrm{d}} \mathbf{r}\ \varepsilon_{\omega}''(\mathbf{r}){\bar{\bar{G}}}_{\omega,n}(\mathbf{r}_1,\mathbf{r}){\bar{\bar{G}}_{\omega,n}^{*}}(\mathbf{r},\mathbf{r}_2)
=\frac{c^2}{\omega^2} \text{Im}\left[ {\bar{\bar{G}}}_{\omega,n}(\mathbf{r}_1,\mathbf{r}_2)\right].
\end{align}
We have defined the scalar product $\braf \mathbf{h}_1 ,\mathbf{h}_2\ketf$ in $L_2(\mathbb{R}^3,\mathbb{C}^3)$, i.e. between two complex vectors which depend on $\mathbf{r}$, as:
\begin{equation}
\braf  \mathbf{h}_1 ,\mathbf{h}_2\ketf:=\int d^3r \ \mathbf{h}_2(\mathbf{r})\cdot \mathbf{h}_1^{\ast}(\mathbf{r}).
\end{equation}
The phase of $\kappa_{\omega,n}(\mathbf{r}_l)$ can be chosen arbitrarily.
The operators $\hat{a}_{\omega,n}(\mathbf{r}_l)$ satisfy the commutation relations
\begin{equation}
\left[\hat{a}_{\omega,n}(\mathbf{r}_l),\hat{a}^{\dagger}_{\omega',n'}(\mathbf{r}_l)\right]=\delta_{nn'}\delta(\omega-\omega').
\end{equation}
Using the above definitions, one can write Eq.\eqref{Hcg} as
\begin{equation}\begin{split}
&\hat{H}=\int_0^{+\infty} d\omega \ \hbar \omega \int d^3r \ \hat{\mathbf{f}}^{\dagger}_{\omega}(\mathbf{r})\hat{\mathbf{f}}_{\omega}(\mathbf{r})+\sum_{l=1}^N \hbar \omega_{eg}\hat{\sigma}_{ee}^{(l)}\\
&-i\hbar\sum_{l=1}^N \hat{\sigma}_{eg}^{(l)} \otimes \int_0^{+\infty} d\omega \sum_n\kappa_{\omega,n}(\mathbf{r}_l)\hat{a}_{\omega,n}(\mathbf{r}_l)+H.c.,
\end{split}
\label{Ha}
\end{equation}
where each emitter interacts with one effective LSP field. 
The operator $\hat{a}_{\omega,n}^{\dagger}(\mathbf{r}_l)$ is associated to the creation of a quantized plasmon by the emitter at the position $\mathbf{r}_l$.

The free plasmonic term includes dark modes, i.e. that are not involved in the coupled dynamics. In the following we construct the dark operators $\hat{\mathbf{d}}_{\omega}(\mathbf{r})$, that will be eliminated from the model.

\subsection{Identification and elimination of the dark modes}

\subsubsection{Hilbert space structure generated by the creation-annihilation operators  $\hat f^{i}(\mathbf{r})$}
\label{Hstruct}

The linear decomposition \eqref{a_def} of the general form
\begin{equation}
\hat{a}=\int d^3r \ \mathbf{h}(\mathbf{r})\cdot \hat{\mathbf{f}}(\mathbf{r})=\sum_{i=1}^{3}\int d^3r\  h^i(\mathbf{r})\hat{f}^i(\mathbf{r})
\end{equation}
with $\mathbf{h}(\mathbf{r})\in L_2(\mathbb{R}^3,\mathbb{C}^3)$ defines a vector space $V_f$, on which we can define a scalar product with the commutator:
\begin{equation}
\braf  \hat{a}_1 ,\hat{a}_2\ketf:= [\hat{a}_{2},\hat{a}^{\dagger}_{1}]=\braf  \mathbf{h}_1 ,\mathbf{h}_2\ketf.
\label{scal_prod}
\end{equation}
Since each element $\hat{a}$ is uniquely represented by the coefficient 
\begin{equation}
h^i(\mathbf{r})=\braf  \hat{f}^i(\mathbf{r}), \hat{a}\ketf=[\hat{a},\hat{f}^{i\dagger}(\mathbf{r})],
\end{equation}
$V_f$ is isomorphic to $L_2(\mathbb{R}^3,\mathbb{C}^3)$ and we use the same notation for the scalar product in \eqref{scal_prod}. The operators $\hat{f}^i(\mathbf{r})$ form an orthonormal basis of the space $V_{\hat f}$.


With the indices of the present problem, each operator $\hat{a}_{\omega,n}(\mathbf{r}_l)$
can be uniquely represented as
\begin{equation}
\braf  \hat{f}_{\omega'}(\mathbf{r}), \hat{a}_{\omega,n}\ketf=[\hat{a}_{\omega,n}(\mathbf{r}_l),\mathbf{\hat{f}}^{\dagger}_{\omega'}(\mathbf{r})]=\mathbf{h}_{\omega,n}(\mathbf{r}_l,\mathbf{r})\delta(\omega-\omega')
\end{equation}
and the commutation relations for two different positions reads
\begin{equation}
[\hat{a}_{\omega,n}(\mathbf{r}_i),\hat{a}_{\omega',n'}^{\dagger}(\mathbf{r}_j)]=\delta_{nn'}\delta(\omega-\omega')\mu^{ij}_{\omega,n},
\label{aa}
\end{equation}
where 
\begin{equation}
\mu^{i,j}_{\omega,n}=
\braf  \mathbf{h}_{\omega,n}(\mathbf{r}_j),\mathbf{h}_{\omega,n}(\mathbf{r}_i)\ketf
\label{mu1}
\end{equation}
is the overlap of the modes labelled by $n$ and $\omega$ excited by the $i^{th}$ and $j^{th}$ QE's.
The operators $\hat{a}_{\omega,n}(\mathbf{r}_l)$ are thus not orthogonal in the QE-index.

\subsubsection{A single emitter}

For a single QE (at the position $\mathbf{r}_1$), we construct the dark operator by subtracting from the field operator $\hat{\mathbf{f}}_{\omega}(\mathbf{r})$ its projection on the subspace generated by the orthogonal set of operators $\hat{a}_{\omega,n}(\mathbf{r}_1)$ in a similar way of a step in a Gram-Schmidt orthogonalization procedure \citep{David,Benjamin}:
\begin{subequations}
\begin{eqnarray}
\hat{\mathbf{d}}_{\omega}(\mathbf{r}) & := & \hat{\mathbf{f}}_{\omega}(\mathbf{r} )- \sum_{n}   \hat{a}_{\omega,n}(\mathbf{r}_1)
 \left[ \hat{a}_{\omega,n}^{\dagger}(\mathbf{r}_1) ,\hat{\mathbf{f}}_{\omega}(\mathbf{r} )  \right]   \\
 & = &\hat{\mathbf{f}}_{\omega}(\mathbf{r} ) - \sum_{n} \hat{a}_{\omega,n}(\mathbf{r}_1) \mathbf{h}_{\omega,n}^{\ast}(\mathbf{r}_1,\mathbf{r}).
\end{eqnarray}
\end{subequations}
The dark operators satisfy the following properties 
\begin{subequations}
\begin{eqnarray}
\left[\hat{a}_{\omega,n}(\mathbf{r}_l),\hat{\mathbf{d}}^{\dagger}_{\omega}(\mathbf{r})\right]&=&\left[\hat{a}_{\omega,n}(\mathbf{r}_l),\hat{\mathbf{d}}_{\omega}(\mathbf{r})\right]=\mathbf{0},\\
\int d^3r\ \hat{\mathbf{f}}^{\dagger}_{\omega}(\mathbf{r})\cdot\hat{\mathbf{f}}_{\omega}(\mathbf{r})&=& \sum_{n} \hat{a}_{\omega,n}^{\dagger}(\mathbf{r}_1) \hat{a}_{\omega,n}(\mathbf{r}_1) \nonumber\\
&&+  \int d^3r\ \hat{\mathbf{d}}_{\omega}^{\dagger}(\mathbf{r}) \cdot \hat{\mathbf{d}}_{\omega}(\mathbf{r}).
\end{eqnarray}
\end{subequations}
Omitting the dark modes that are not populated by the interaction with the emitter, we derive the reduced Hamiltonian from the above decomposition as anticipated in \eqref{Hred}:
\begin{equation}\begin{split}
\hat{H}&=\int_0^{+\infty} d\omega \ \hbar \omega\sum_{n} \hat{a}_{\omega,n}^{\dagger}(\mathbf{r}_1) \hat{a}_{\omega,n}(\mathbf{r}_1)+\hbar \omega_{eg}\hat{\sigma}_{ee}^{(1)}\\ 
&-\hat{\sigma}_{eg}^{(1)} \otimes \int_0^{+\infty} d\omega\ \sum_{n}\kappa_{\omega,n}(\mathbf{r}_1)\hat{a}_{\omega,n}(\mathbf{r}_1)+H.c.
\end{split}
\label{Hred1}
\end{equation}
We can remark that the above construction of bright and dark modes can be done also with global mode operators, i.e. without separating them with respect to the index $n$.

Using the concepts defined in Sec. \ref{Hstruct}, we can reinterpret the construction as follows: The bright operators $ \hat{a}_{\omega,n}(\mathbf{r}_1)$ span a subspace $V_{\hat a}$ of $V_{\hat f}$, of which they are an orthonormal basis.
The dark mode operators are in the orthogonal complement of  $V_{\hat a}$.

\subsubsection{Many emitters: Orthonormalization of the operators $\hat{a}_{\omega,n}(\mathbf{r}_i)$}

The generalization of this procedure is not direct for $N>1$ emitters since the set of operators $\hat{a}_{\omega,n}(\mathbf{r}_i)$ is not orthogonal.
To solve this problem, 
we construct a set of bright operators that are mutually orthonormal by taking suitable linear combinations of the $\hat{a}_{\omega,n}(\mathbf{r}_i)$:
\begin{equation}
\hat{b}_{\omega,n}^{(j)}= \sum_{i=1}^{N} \beta^{j,i}_{\omega,n}\hat{a}_{\omega,n}(\mathbf{r}_i), \quad j=1,\cdots,N_{\text{ind}},
\label{nuovi}
\end{equation}
where $N_{\text{ind}}\le N$ is the number of linearly independent operators $\hat{a}_{\omega,n}(\mathbf{r}_i)$ and the coefficients $\beta^{j,i}_{\omega,n}$ are chosen such that the new operators satisfy the orthonormality condition 
\begin{subequations}
\begin{align}
\left[\hat{b}_{\omega,n}^{(i)},\hat{b}_{\omega',n'}^{(j)\dagger} \right]&=\delta_{nn'}\delta(\omega-\omega')\delta_{ij},\\
\left[\hat{b}_{\omega,n}^{(i)},\hat{b}_{\omega',n'}^{(j)}\right]&=0.
\end{align}
\label{orthcond}
\end{subequations}
They can be constructed by the Gram-Schmidt method \cite{David} or by other orthonormalization procedures, as we discuss below.
 
In order to implement the orthonormalization procedure, we have to identify first the number of linearly independent field $\hat{a}_{\omega,n}(\mathbf{r}_i)$. In \cite{David}, it has been shown that if $|\mu^{i,j}_{\omega,n}|=1$ with $i>j$, multiplying the relation \eqref{aa} by $\mu^{i,j\ast}_{\omega',n'}$ leads to
\begin{equation}
\hat{a}_{\omega,n}(\mathbf{r}_i)=\mu^{i,j}_{\omega,n}\hat{a}_{\omega,n}(\mathbf{r}_j).
\end{equation}
In this way we can identify pair of linearly dependent field operators. There is also the possibility that one operator can be expressed as linear combination of two or more other ones. To identify the number $N_{\text{ind}}$, we define for the $N$ LSP effective field operators $\{\hat{a}_{\omega,n}(\mathbf{r}_l)\}$, the $N\times N$ \textit{overlap matrix} $M_{\omega,n}$ (see Appendix A):
\begin{equation}
M_{\omega,n}:=\begin{pmatrix}
1 & \mu^{2,1}_{\omega,n} & \cdots & \mu^{N,1}_{\omega,n}\\
 \mu^{1,2}_{\omega,n} & \ddots & \ddots & \vdots\\
\vdots & \ddots & \ddots & \mu^{N,N-1}_{\omega,n} \\
\mu^{1,N}_{\omega,n}  & \cdots & \mu^{N-1,N}_{\omega,n} & 1
\end{pmatrix}.
\label{overlapmatrix}
\end{equation}
The overlap matrix is hermitian, so we can diagonalize it through a $N \times N$ unitary matrix $T_{\omega,n}$
\begin{equation}
\begin{split}
&T^{\dagger}_{\omega,n} M_{\omega,n}T_{\omega,n}\equiv D_{\omega,n} \\
&=\text{diag}\left(\lambda_{1,\omega,n},\cdots,\lambda_{N_{\text{ind}},\omega,n},0,\cdots,0\right).
\end{split}
\label{diagM}
\end{equation}
We can prove that the number of non-zero eigenvalues is the number of linearly independent operators (see Appendix A). Implementing, for example, the Gram-Schmidt orthogonalization procedure, we can obtain an orthonormal set of $N_{\text{ind}}$ bright operators in terms of $\{\hat{a}_{\omega,n}(\mathbf{r}_i)\}$ with $i=1,\cdots,N_{\text{ind}}$\cite{David}. To obtain the analytical expressions of the coupling constants, we have to derive, from the knowledge of the overlap matrix, the expressions of the operators $\{\hat{a}_{\omega,n}(\mathbf{r}_{N_{\text{ind}}+1}),\cdots,\hat{a}_{\omega,n}(\mathbf{r}_{N})\}$ in terms of $\{\hat{a}_{\omega,n}(\mathbf{r}_1),\cdots,\hat{a}_{\omega,n}(\mathbf{r}_{N_{\text{ind}}})\}$, and substitute them in the interaction Hamiltonian in terms of the new orthonormal ones.

\subsubsection{L\"{o}wdin orthonormalization of the operators $\hat{a}_{\omega,n}(\mathbf{r}_l)$ and Singular Value Decomposition}

The Gram-Schmidt method is a \textit{sequential} technique of orthonormalization, i.e. the $m^{\text{th}}$ new operator can be obtained after having derived the $m-1$ ones. In this paper we generalize the L\"{o}wdin's canonical orthonormalization \citep{Lowdin} to the case in which $N$ initial operators are not necessarily linearly independent. Moreover, this technique has the advantage that it gives us a \textit{global} algorithm, i.e. it considers simultaneously all the vectors to be orthonormalized. The fact formulated in Section \ref{Hstruct} that the commutators of boson creation-annihilation operators can be interpreted as scalar products allows us to apply directly the results of the  L\"owdin orthogonalization to the construction of the bright and dark boson operators.  

Defining the following $N \times N$ matrix
\begin{equation}
D_{-1/2;\omega,n}:= \text{diag}\left(\lambda_{1,\omega,n}^{-1/2},\cdots,\lambda_{N_{\text{ind}},\omega,n}^{-1/2},0,...,0\right),
\end{equation}
we implement the L\"{o}wdin's canonical orthonormalization, obtaining a new set of bright operators $\{\hat{b}_{\omega,n}^{(j)}\}$
\begin{equation}
B_{\omega,n}=A_{\omega,n}T_{\omega,n}D_{-1/2;\omega,n},
\label{Lm}
\end{equation}
through the one row arrays
\begin{equation}
B_{\omega,n}:=\left[\hat{b}_{\omega,n}^{(1)},\cdots , \hat{b}_{\omega,n}^{(N_{\text{ind}})}, 0,\cdots ,0\right]
\end{equation}
\begin{equation}
A_{\omega,n}:=\left[\hat{a}_{\omega,n}(\mathbf{r}_1),\cdots , \hat{a}_{\omega,n}(\mathbf{r}_{N})\right].
\end{equation}
Each L\"{o}wdin operator $\hat{b}_{\omega,n}^{(j)}$ can be expressed in terms of the old ones as follows
\begin{equation}
\hat{b}_{\omega,n}^{(j)}=
\begin{cases}
\lambda_j^{-1/2}\sum_{i=1}^N T^{i,j}_{\omega,n}\hat{a}_{\omega,n}(\mathbf{r}_i) & j \in [1,N_{\text{ind}}]\\
0 & j \in [N_{\text{ind}}+1,N].
\end{cases}
\label{lsingoli}
\end{equation}

\noindent We can express the old operators $\hat{a}_{\omega,n}(\mathbf{r}_i)$ in terms of L\"{o}wdin's operators as a step of the \textit{Singular Value Decomposition} of $B_{\omega,n}$ \cite{Lowdin} (see Appendix B)
\begin{equation}
A_{\omega,n}=B_{\omega,n}D^{1/2}_{\omega,n}T^{\dagger}_{\omega,n}
\end{equation}
by inverting \eqref{lsingoli}:
\begin{equation}
\hat{a}_{\omega,n}(\mathbf{r}_i)=\sum_{j=1}^{N_{\text{ind}}}\lambda^{1/2}_{j,\omega,n}T^{i,j\ast}_{\omega,n}\hat{b}_{\omega,n}^{(j)} \quad i\in [1,N].
\label{asingolil}
\end{equation}
If the $N$ operators $\{\hat{a}_{\omega,n}(\mathbf{r}_i)\}$ are linearly independent, in Eq. \eqref{diagM} no eigenvalue is zero. According to the expressions \eqref{Lm} and \eqref{lsingoli}, L\"{o}wdin's method gives us $N$ orthonormal mode-selective bright LSP operators. If $N_{\text{ind}}<N$, at least one of the eigenvalues is zero and the new set in \eqref{lsingoli} shows a reduced dimension.
The expression \eqref{asingolil} can be directly substituted in the interaction term of \eqref{Ha} [without knowing the expressions of the operators $\{\hat{a}_{\omega,n}(\mathbf{r}_{N_{\text{ind}+1}}),\cdots,\hat{a}_{\omega,n}(\mathbf{r}_{N})\}$]:
\begin{equation}
\hat{H}_{\text{int}}=-i\hbar\sum_{i=1}^N \hat{\sigma}_{eg}^{(i)}\otimes\int_0^{+\infty} d\omega\sum_n\sum_{j=1}^{N_{\text{ind}}} \kappa^{i,j}_{\omega,n}\hat{b}_{\omega,n}^{(j)}+h.c.
\label{Hint_final}
\end{equation}
The coupling constants of each bright field $\hat{b}_{\omega,n}^{(j)}$ with each emitter are expressed in terms of the eigenvalues and of the eigenvectors of the overlap matrix $M_{\omega,n}$ in the following compact form
\begin{equation}
\kappa^{i,j}_{\omega,n}=\kappa_{\omega,n}(\mathbf{r}_i)\lambda_{j,\omega,n}^{1/2}T^{i,j\ast}_{\omega,n}.
\label{accopLow}
\end{equation}
Identifying \eqref{lsingoli} with \eqref{nuovi}, we obtain:
\begin{equation}
\beta^{j,i}_{\omega,n}=\lambda_j^{-1/2} T^{i,j}_{\omega,n}.
\label{coef_beta}
\end{equation}

\subsubsection{Construction of the dark operators}
The bright operators are the only ones that appear in the interaction term of the Hamiltonian \eqref{Hint_final}. The final step is to express the free term of the Hamiltonian in terms of the bright operators and another set of operators, the {\it dark operators}, which are not coupled to the emitters and that are orthogonal to the bright operators. Once we have the set of bright operators  $\{  \hat b_{\omega,n}^{(j)}  \}$  that are orthonormal to each other, the construction of the dark operators takes the same general form as what we did for the single emitter case.
We define the dark operators as follows
\begin{equation}
\hat{\mathbf{d}}_{\omega}(\mathbf{r})=\hat{\mathbf{f}}_{\omega}(\mathbf{r})-\sum_n\sum_{i=1}^{N_{\text{ind}}}\hat{b}_{\omega,n}^{(i)}[\hat{\mathbf{f}}_{\omega}(\mathbf{r}),\hat{b}_{\omega,n}^{(i)\dagger}].
\label{darkb}
\end{equation}
The bright and dark  operators satisfy the following properties
\begin{eqnarray}
&(i)&\  [ \hat b^{(i)\dag}_{\omega,n},\hat{\mathbf{d}}_{\omega}(\mathbf{r})]  =  \mathbf{0}, \quad  [ \hat b_{\omega,n}^{(i)},\hat{\mathbf{d}}_{\omega}(\mathbf{r})] =\mathbf{0}\qquad  \\
\label{result} \nonumber
&(ii)&\ H_0= \int_0^\infty d\omega~\hbar\omega\int d^3r\ \hat{\mathbf{f}}_\omega^{\dag}(\mathbf{r})\cdot \hat{\mathbf{f}}_\omega(\mathbf{r})   \\ 
 & & \qquad= \int_0^\infty d\omega~\hbar\omega \sum_{n}\sum_{i=1}^{N_{\text{ind}}} \hat b_{\omega,n}^{(i)\dag} \hat b_{\omega,n}^{(i)}\nonumber\\
 & &\qquad\qquad+ \int_0^\infty d\omega~\hbar\omega \int d^3r\ \hat{\mathbf{d}}_\omega^{\dag}(\mathbf{r})  \cdot \hat{\mathbf{d}}_\omega(\mathbf{r}).
\end{eqnarray}
The final mode-selective continuous microscopic model can be written as follows
\begin{equation}
\begin{split}
&\hat H=\int_0^\infty d\omega~\hbar\omega \sum_{n}\sum_{i=1}^{N_{\text{ind}}} \hat b_{\omega,n}^{(i)\dag} \hat b_{\omega,n}^{(i)}
+\sum_{i=1}^{N}\hbar \omega_{eg} \hat{\sigma}_{ee}^{(i)}\\
& -i\hbar\sum_{i=1}^N \hat{\sigma}_{eg}^{(i)}\otimes\int_0^{+\infty} d\omega\sum_n\sum_{j=1}^{N_{\text{ind}}} \kappa^{i,j}_{\omega,n}\hat{b}_{\omega,n}^{(j)}+h.c.
\end{split}
\end{equation}
We remark that, in practice, in order to write the effective Hamiltonian we only need to determine the coupling constants $ \kappa^{i,j}_{\omega,n}$. The coefficients  $\beta^{j,i}_{\omega,n}$ \eqref{coef_beta} are only needed for the theoretical justification of the separation of the bright and the dark modes.

The L\"{o}wdin's method provides simple compact formulas using an algorithm, which is more stable than the Gram-Schmidt algorithm in numerical implementations, allowing in principle the treatment of a large number of emitters.

\subsubsection{L\"{o}wdin orthonormalization for $N=2$}
\noindent For the two LSP effective field operators $\hat{a}_{\omega,n}(\mathbf{r}_1)$ and $\hat{a}_{\omega,n}(\mathbf{r}_2)$, the overlap matrix
\begin{equation}
M_{\omega,n}=
\begin{pmatrix}
1 & \mu^{21}_{\omega,n} \\ \\
\mu^{12}_{\omega,n} & 1
\end{pmatrix}
\label{overlapmatrix2}
\end{equation}
can be diagonalized by a unitary matrix
\begin{equation}
T_{\omega,n}=\begin{pmatrix}
\dfrac{1}{\sqrt{2}}\dfrac{\mu^{21}_{\omega,n}}{|\mu^{12}_{\omega,n}|} \ & \ -\dfrac{1}{\sqrt{2}}\dfrac{\mu^{21}_{\omega,n}}{|\mu^{12}_{\omega,n}|} \\ \\
\dfrac{1}{\sqrt{2}} \ & \ \dfrac{1}{\sqrt{2}} 
\end{pmatrix}
\label{T}
\end{equation}
such that
\begin{equation}\begin{split}
T^{\dagger}_{\omega,n}&M_{\omega,n}T_{\omega,n}=D_{\omega,n}:=\text{diag}\left(\lambda_{1,\omega,n},\lambda_{2,\omega,n}\right)
\end{split}
\end{equation}
with the eigenvalues
\begin{equation}
\lambda_{1,\omega,n}=1+|\mu^{12}_{\omega,n}|,\quad
\lambda_{2,\omega,n}=1-|\mu^{12}_{\omega,n}|.
\end{equation}
We define the matrix $D_{-1/2;\omega,n}$ that, if $N_{\text{ind}}=N=2$, has the following form
\begin{equation}
D_{-1/2;\omega,n}=\text{diag}\left(\lambda_{1,\omega,n}^{-1/2},\lambda_{2,\omega,n}^{-1/2}\right),
\label{Dind}
\end{equation}
otherwise
\begin{equation}
D_{-1/2;\omega,n}=\text{diag}\left(\lambda_{1,\omega,n}^{-1/2},0\right).
\label{Ddip}
\end{equation}
Implementing the L\"{o}wdin's canonical orthonormalization, we obtain a new set of bright operators
\begin{subequations}
\begin{equation}\begin{split}
\hat{b}^{(1)}_{\omega,n}&=\dfrac{1}{\sqrt{2}}\dfrac{1}{\sqrt{1+|\mu^{12}_{\omega,n}|}}\left[\hat{a}_{\omega,n}(\mathbf{r}_2)+\dfrac{\mu^{21}_{\omega,n}}{|\mu^{12}_{\omega,n}|}
\hat{a}_{\omega,n}(\mathbf{r}_1)\right]
\end{split}
\label{b1}
\end{equation}
\begin{equation}
\begin{split}
\hat{b}^{(2)}_{\omega,n}&=\dfrac{1}{\sqrt{2}}\dfrac{1}{\sqrt{1-|\mu^{12}_{\omega,n}|}}\left[\hat{a}_{\omega,n}(\mathbf{r}_2)-\dfrac{\mu^{21}_{\omega,n}}{|\mu^{12}_{\omega,n}|}\hat{a}_{\omega,n}(\mathbf{r}_1)\right]
\end{split}
\label{b2}
\end{equation}
\end{subequations}
satisfying the orthonormality condition \eqref{orthcond}.
If the two operators are linearly independent, $|\mu^{12}_{\omega,n}|\neq 1$ and $\lambda_{2,\omega,n}^{-1/2}\neq 0$,
the L\"{o}wdin's method gives us two new orthonormal mode-selective LSP operators. If $N_{\text{ind}}=1$, $|\mu^{12}_{\omega,n}|=1$ \citep{David}, the second eigenvalue is zero:
the technique returns a single new operator.

Expressing the old effective operators $\hat{a}_{\omega,n}(\mathbf{r}_i)$ in terms of the new one(s)
and substituting them in the interaction term of eq. \eqref{Ha}, we obtain the coupling constants
\begin{subequations}
\begin{equation}
\kappa^{11}_{\omega,n}=\kappa_{\omega,n}(\mathbf{r}_1)\dfrac{\mu^{12}_{\omega,n}}{|\mu^{12}_{\omega,n}|}\dfrac{1}{\sqrt{2}}\sqrt{1+|\mu^{12}_{\omega,n}|},
\end{equation}
\begin{equation}
\kappa^{12}_{\omega,n}=-\kappa_{\omega,n}(\mathbf{r}_1)\dfrac{\mu^{21}_{\omega,n}}{|\mu^{12}_{\omega,n}|}\dfrac{1}{\sqrt{2}}\sqrt{1-|\mu^{12}_{\omega,n}|},
\label{copl12}
\end{equation}
\begin{equation}
\kappa^{21}_{\omega,n}=\kappa_{\omega,n}(\mathbf{r}_2)\dfrac{1}{\sqrt{2}}\sqrt{1+|\mu^{12}_{\omega,n}|},
\end{equation}
\begin{equation}
\kappa^{22}_{\omega,n}=\kappa_{\omega,n}(\mathbf{r}_2)\dfrac{1}{\sqrt{2}}\sqrt{1-|\mu^{12}_{\omega,n}|}.
\label{copl22}
\end{equation}
\end{subequations}
We can see that if $|\mu^{12}_{\omega,n}|=1$, i.e. the two original non orthogonal operators are not linearly independent, the theory can be formulated only in terms of the first operator $\hat{b}^{(1)}_{\omega,n}$ and the coupling constants linking the two QEs to the second field, $\kappa^{12}_{\omega,n}$ and $\kappa^{22}_{\omega,n}$, are automatically zero, as shown by \eqref{copl12} and \eqref{copl22}.

\subsection{Discrete effective model}
A next step consists of getting rid of the dependence on the continuous parameter $\omega$ and of truncating the infinite Hilbert space of the system.

\noindent It has been proved in \cite{David} that the QEs-LSP coupling constants can be very well approximated by
a Lorentzian function for each plasmonic mode:
\begin{equation}
\kappa_{\omega,n}(\mathbf{r}_i)=g_n(\mathbf{r}_i)L_{n}(\omega)=
\sqrt{\dfrac{\gamma_n}{2\pi}}\dfrac{g_n(\mathbf{r}_i)}{\omega-\omega_n+i\gamma_n}.
\label{lor}
\end{equation}
where $L_{n}(\omega)$ indicates the Lorentzian function, 
$\gamma_n$ is its half-width, i.e. the lossy rate of the $n^{th}$ mode,
and $\omega_n$ is its resonance frequency. 
Exploiting this coupling shape \eqref{lor}, it is possible to get rid of the dependence on the continuous
parameter $\omega$ by defining $N_{\text{ind}}$ effective operators $\hat{b}^{(i)}_{n}$ \cite{David}:
\begin{equation}
\hat{b}^{(i)}_n=\int_0^{+\infty}d\omega L_{n}(\omega)\hat{b}^{(i)}_{\omega,n}
\end{equation}
allowing us to derive a discrete effective model describing the interaction of each QE with bosonic plasmonic resonances
\begin{equation}\begin{split}
&\hat{H}_d=\sum_n \sum_{i=1}^{N_{\text{ind}}}\hbar(\omega_n-i\gamma_n)\hat{b}^{(i)\dagger}_n\hat{b}^{(i)}_n+\sum_{i=1}^{N}\hbar\omega_{eg}\hat{\sigma}_{ee}^{(i)}\\
&-i\hbar\sum_{i=1}^N \hat{\sigma}_{eg}^{(i)}\otimes \sum_n\sum_{j=1}^{N_{\text{ind}}}[g^{ij}_n\hat{b}^{(j)}_n-h.c.]
\end{split}
\label{effectdiscrmodel1}
\end{equation}
with
\begin{equation}
g^{ij}_n=g_n(\mathbf{r}_i)\lambda_{j,\omega_n,n}^{1/2}T^{i,j\ast}_{\omega_n,n},
\end{equation}
where the eigenvalues and the eigenvectors of the overlap matrix is considered at the resonance frequency of the mode $n$: $M_{\omega_n,n}$.

\begin{figure}[!h]
\begin{center}
\includegraphics[scale=0.2]{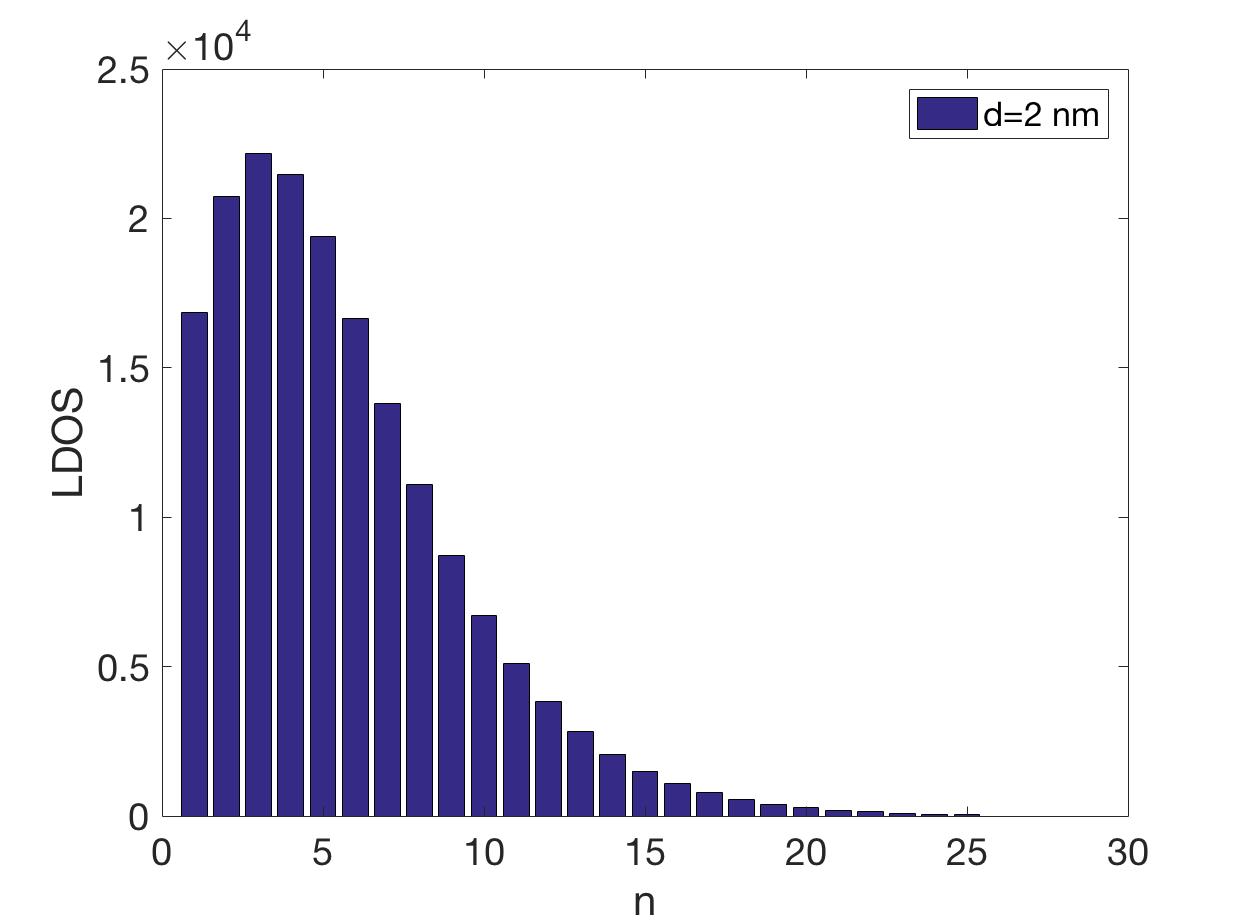}
\includegraphics[scale=0.2]{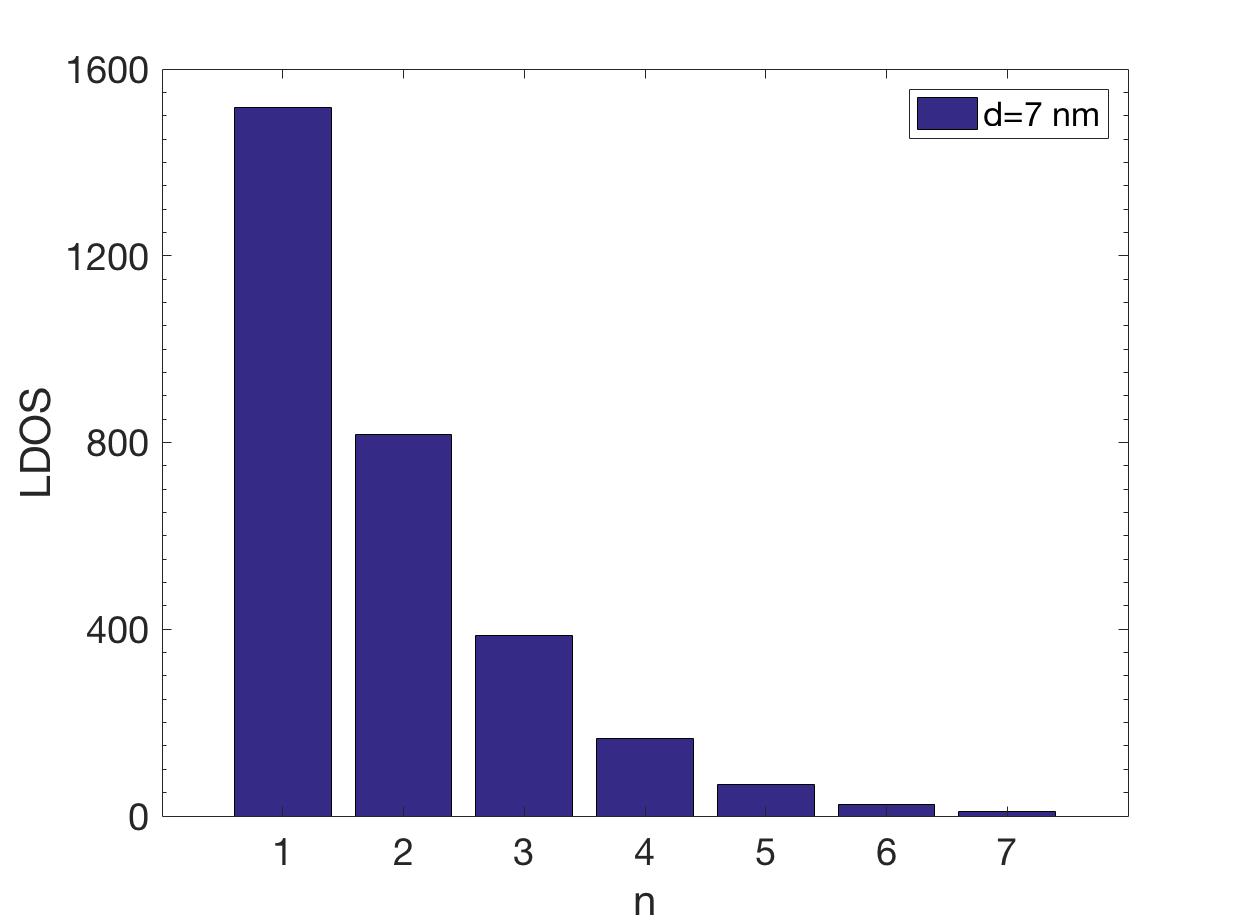}
\end{center}
\caption{Local density of states associated to the different excited LSPP modes, for a radially polarized emitter placed at a distance $d=2$ nm (upper frame) and $d=7$ nm (lower frame), respectively, from the metallic surface.}
\label{fig:fig1}
\end{figure}

The effective model \eqref{effectdiscrmodel1} can be truncated since not all the LSP modes play a role in the coupled
dynamics. Indeed, the LDOS of some modes at certain distances from the MNPs surface, i.e. at the QEs-positions, is negligible. Two example of this are plotted in Fig. \ref{fig:fig1} 

Consequently we truncate the model keeping
$n \in [0,n']$,
where $n'$ is the harmonic index associated to the last LSP mode effectively involved in the coupled dynamics. For $n>n'$ the LDOS at the QE-position is approximately zero. 

\noindent In the case of $N=2$, we assume the single excitation subspace spanned by the following basis
\begin{equation}\begin{split}
&\{|e,g;[\mathbf{0}]\rangle , |[\mathbf{g}];1_1, 0 \rangle ,\cdots, |[\mathbf{g}]; 1_{n'},0 \rangle,\\
&|[\mathbf{g}];0,1_1 \rangle ,\cdots, |[\mathbf{g}];0, 1_{n'} \rangle, |g,e;[\mathbf{0}]\rangle \}
\end{split}
\label{basisn'}
\end{equation}
leading to the effective matrix Hamiltonian in this basis \eqref{basisn'}
\begin{equation}\begin{split}
&\hat{H}_{d}=\\ 
&\begin{pmatrix}\begin{array}{c| c  c  c | c  c  c | c}
0 & \cdots & g^{11}_n & \cdots & \cdots & g^{12}_n & \cdots & 0 \\ 
\hline
\vdots & \ddots & & & 0 & \cdots & 0 & \vdots\\
g^{11 \ast}_n & & \Delta_n-i\dfrac{\gamma_n}{2} & & \vdots & & \vdots & g^{21 \ast}_n \\
\vdots & & & \ddots & 0 & \cdots & 0 & \vdots\\
\hline
\vdots & 0 & \cdots & 0 & \ddots & &  & \vdots\\
g^{12 \ast}_n & \vdots & &  \vdots & & \Delta_n-i\dfrac{\gamma_n}{2} & & g^{22 \ast}_n \\
\vdots & 0 & \cdots & 0 &  &  & \ddots & \vdots\\
\hline
0 & \cdots & g^{21}_n & \cdots & \cdots & g^{22}_n & \cdots & 0
\end{array}
\end{pmatrix}.
\end{split}
\label{Heff2}
\end{equation} 

\section{STIRAP PROCESS} \label{III}
\subsection{Presentation}
\begin{figure}[!h]
\begin{center}
\includegraphics[scale=0.15]{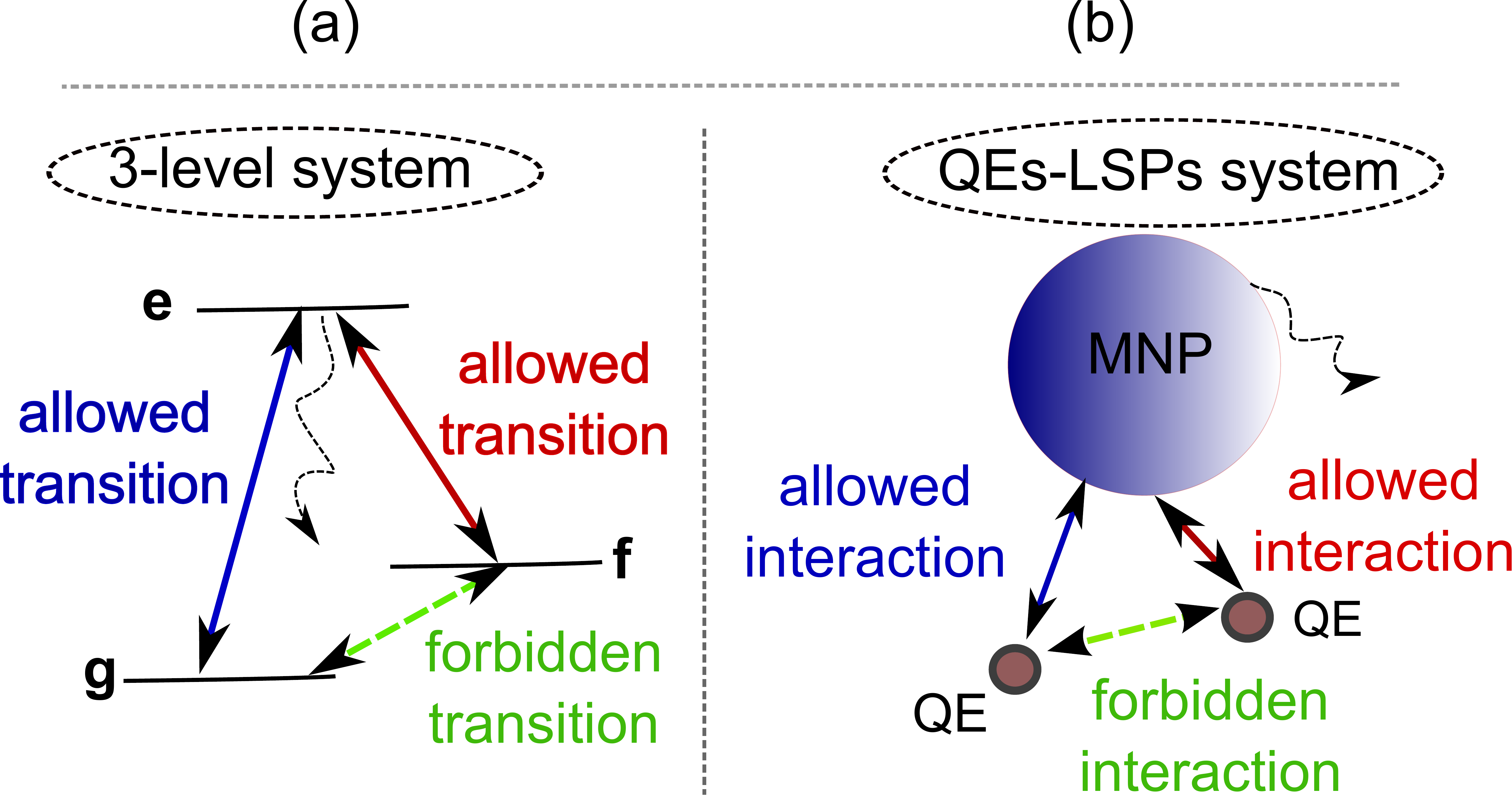}
\caption{(a): Allowed and forbidden transitions in a standard 3-level system in which STIRAP is usually applied. (b): Allowed and forbidden interactions in the analysed QEs-LSPs system.}
\label{fig:fig3}
\end{center}
\end{figure}
We analyze the possible quantum channel that exploits the plasmonic coupling of each emitter with the MNP, through which two QEs can exchange their population. 
We consider the STIRAP process between two atoms. Figure \ref{fig:fig3} shows a qualitative sketch of the analogy between the 3-level system (Fig.
\ref{fig:fig3}a) in which STIRAP is usually applied and the present system (Fig. \ref{fig:fig3}b). Like the two states $|1\rangle$ and $|3\rangle$ involved in the transfer in Fig. \ref{fig:fig3}a, the two emitters in Fig. \ref{fig:fig3}b are assumed not directly coupled. The MNP, that can communicate with both emitters, play the same role of the intermediate lossy excited state $|2\rangle$ but in a tensored product representation for the two atoms and the plasmonic dressing field.

For two three-state QE's we have to introduce two additional states in the basis \eqref{basisn'}:
$|f,g;[\mathbf{0}]\rangle$, $|g,f;[\mathbf{0}]\rangle$.
We assume that the state $|\Psi(t)\rangle$ of the global hybrid system is initially prepared as follows 
\begin{equation}
|\Psi(t\to-\infty)\rangle = |f,g;[\mathbf{0}]\rangle.
\end{equation}
We seek a control setup to obtain a final population exchange of the two QEs, i.e. $|\Psi(t\to+\infty)\rangle =|g,f;[\mathbf{0}]\rangle$.

\begin{figure}[!h]
\begin{center}
\includegraphics[scale=0.7]{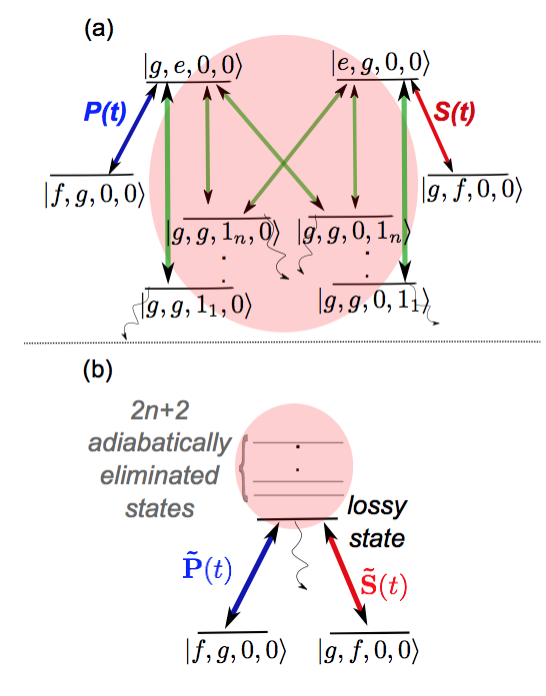}
\end{center}
\caption{(a) Multilevel scheme of the allowed transitions characterizing the laser controlled system composed by two QEs and the MNP. $P(t)$ and $S(t)$ are the Rabi frequencies associated to the interactions with the pump and the Stokes laser, respectively. (b) Effective 3-state system obtained by adiabatic elimination. The Rabi frequencies $\tilde{P}(t)$ and $\tilde{S}(t)$ quantify the coupling strength of the states $|f,g,0,0\rangle$ and $|g,f,0,0\rangle$ respectively with only one of the $2n+2$ intermediate lossy states.}
\label{fig:fig4}
\end{figure}
\noindent We consider the STIRAP configuration with the pump and Stokes pulses, as shown in Fig. \ref{fig:fig4}(a):
\begin{equation}\begin{split}
\hat{H}_d=&\sum_n \sum_{i=1}^{N_{\text{ind}}}\hbar(\omega_n-i\gamma_n)\hat{b}^{(i)\dagger}_n\hat{b}^{(i)}_n
+\sum_{i=1}^{N}\hbar\bigl(\omega_{eg}\hat{\sigma}_{ee}^{(i)}+\omega_{fg}\hat{\sigma}_{ff}^{(i)}\bigr)\\
&-i\hbar\sum_{i=1}^N \hat{\sigma}_{eg}^{(i)}\otimes \sum_n\sum_{j=1}^{N_{\text{ind}}}[g^{ij}_n\hat{b}^{(j)}_n-h.c.]\\
&+\hbar P(t) e^{-i\omega_P t}\hat{\sigma}_{ef}^{(1)}+\hbar S(t) e^{-i\omega_S t}\hat{\sigma}_{ef}^{(2)}+h.c.
\end{split}
\label{effectdiscrmodel}
\end{equation}
In the resonant case, $\omega_P=\omega_S=\omega_{ef}$, the complete matrix hamiltonian reads
\begin{footnotesize}
\begin{equation}\begin{split}
&H=\\
&\begin{pmatrix}\begin{array}{c c| c  c  c | c  c  c | c c}
0 & P(t) & \cdots & 0 & \cdots & \cdots & 0 & \cdots & 0 & 0 \\
P(t) & 0 & \cdots & g^{11}_n & \cdots & \cdots & g^{12}_n & \cdots & 0 & 0 \\ 
\hline
\vdots & \vdots & \ddots & & & 0 & \cdots & 0 & \vdots & \vdots \\
0 & g^{11 \ast}_n & & \Delta_n-i\dfrac{\gamma_n}{2} & & \vdots & & \vdots & g^{21 \ast}_n & 0\\
\vdots & \vdots & & & \ddots & 0 & \cdots & 0 & \vdots & \vdots \\
\hline
\vdots & \vdots & 0 & \cdots & 0 & \ddots & &  & \vdots & \vdots\\
0 & g^{12 \ast}_n & \vdots & &  \vdots & & \Delta_n-i\dfrac{\gamma_n}{2} & & g^{22 \ast}_n & 0\\
\vdots & \vdots & 0 & \cdots & 0 &  &  & \ddots & \vdots & \vdots \\
\hline
0 & 0 & \cdots & g^{21}_n & \cdots & \cdots & g^{22}_n & \cdots & 0 & S(t)\\
0 & 0 & \cdots & 0 & \cdots & \cdots & 0 & \cdots & S(t) & 0
\end{array}
\end{pmatrix}.
\label{Heff2laser}
\end{split}
\end{equation}
\end{footnotesize}
from the resonant transformation
\begin{equation}
H=U^{\dagger}H_d U-i\hbar U^{\dagger}\partial_t U-\omega_{eg}\openone_{n'}
\end{equation}
with
\begin{equation}
U=\text{diag}(e^{i\omega_P t}, 1, \cdots, 1, e^{i\omega_S t}).
\end{equation}
The STIRAP process can be interpreted in the basis
\begin{equation}
\{|f,g;[\mathbf{0}]\rangle ,|\Phi_1\rangle , |\Phi_2 \rangle ,\cdots, |\Phi_{2n+2} \rangle, |g,f;[\mathbf{0}]\rangle\},
\label{effbasis}
\end{equation}
in which \eqref{Heff2laser} has a $(2n+1)\times(2n+1)$ diagonal central block.
After adiabatic eliminations, the multilevel system can be reduced to an effective 3-state subspace spanned by 
\begin{equation}
\{ |f,g;[\mathbf{0}\rangle, |\Phi_m \rangle, |g,f;[\mathbf{0}]\rangle,
\end{equation}
where $|\Phi_m \rangle$ (with $m \in [1,2n+2]$) plays the same role of the excited state in the standard STIRAP \citep{Benjamin} (see Fig. \ref{fig:fig4}(b)).

\subsection{Dependence of the STIRAP process on the angular distance of the QEs}
\noindent We analyze the dependence of the STIRAP process on the angular distance of the QEs for a metallic sphere of radius $R=8$ nm, characterized by a Drude dielectric function $\epsilon_m(\omega)$ \citep{Zubairy}, and Gaussian laser pulses
\begin{equation}
P(t)=\Omega_0e^{-[(t-\tau)/T]^2} \qquad
S(t)=\Omega_0e^{-[(t+\tau)/T]^2},
\end{equation}
where $2\tau$ is the delay between the two pulses, $T$ is the pulse width and $\Omega_0$ is the peak Rabi frequency. 

\begin{figure}[!h]
\begin{center}
\includegraphics[scale=0.68]{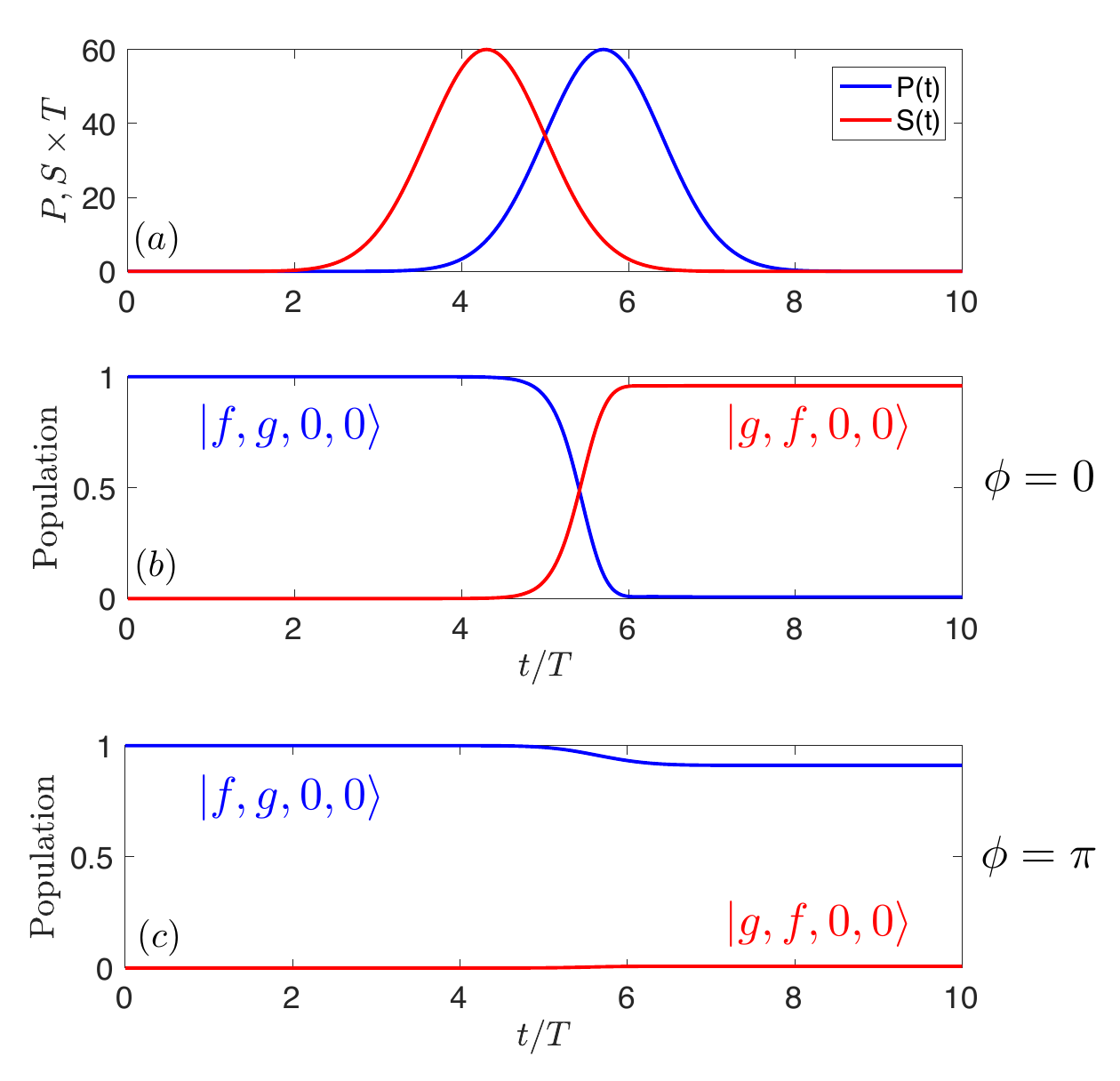}
\end{center}
\caption{Dynamics of the population transfer from $|f,g,0,0\rangle$ to  $|g,f,0,0\rangle$ for $\phi=0$ (middle frame) and $\phi=\pi$ (lower frame) for a MNP of $R=8$ nm interacting with two dipoles of 10 D, one positionned at $2$ nm and the other one at $4$ nm from the metallic surface, and Gaussian profile of the two delayed and counterintuitively ordered laser pulses (upper frame). The area of each pulse is $\Omega_0T=60$ and the delay is $\tau/T=0.7$, where $T\simeq 10$ ns.}
\label{fig:fig5}
\end{figure}

We define $\phi$ as the angle between the position vectors of the two QEs.
We anticipate favorable configurations with an efficient transfer between the two QEs mediated by plasmons, when they satisfy an overlap $|\mu^{12}|\simeq 1$, i.e. for the cases in which the two emitters are aligned in the same side ($\phi=0$) or in the two opposite ones of the sphere ($\phi=\pi$),
almost at the same distance from the metallic surface. Numerical simulations are presented in Fig. \ref{fig:fig5}, for QEs placed very close to the surface ($2$ nm and $4$ nm respectively).
We have numerically found that a STIRAP process can be well implemented for $\phi=0$ (see Fig. \ref{fig:fig5}(b)) since more than 90\% of the population can be transferred from the initial state $|f,g;[\mathbf{0}]\rangle$ to the target state $|g,f;[\mathbf{0}]\rangle$ with a negligible plasmonic population during all the dynamics (giving to the plasmon polaritons the role of a dark subsystem). 
On the other hand, if the QEs are aligned in the opposite sides of the MNP ($\phi=\pi$), the
population is blocked in the initial state (see Fig. \ref{fig:fig5}(c)) \citep{Benjamin}.

\begin{figure}[!h]
\begin{center}
\includegraphics[scale=0.3]{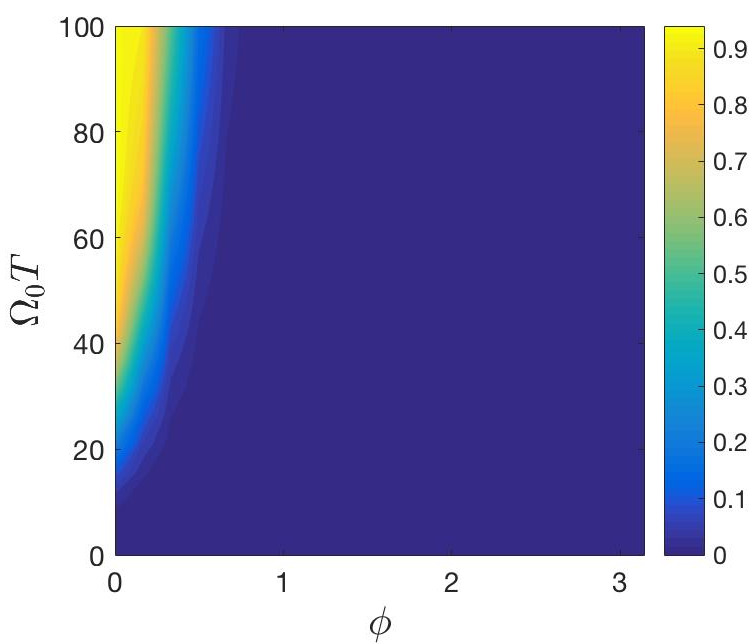}
\end{center}
\caption{Contour plot of the transfer efficiency $|\langle g,f,0,0|\Psi(t=+\infty)\rangle |^2$ as a function of the area $\Omega_0T$ of the laser pulses and of the angle $\phi \in [0,\pi]$ between the position vectors of the two emitters from the center of the MNP.}
\label{fig:fig6}
\end{figure}

We have numerically investigated the dependence of the transfer efficiency on the angular distance $\phi$ in Fig. \ref{fig:fig6}. We found that, under adiabatic conditions,
a good population transfer (not less than 70 \%) can be obtained for very small angles up to $\pi/15$. For larger angles, Fig. \ref{fig:fig6} shows a negligible population transfer.
This shows that the blockade takes place already for a quite small angle.

\subsection{Dependence of the STIRAP process on the number of plasmonic modes}
For a distance between the QEs and the metallic surface of $2$ or $4$ nm, we have considered the first 25 plasmonic modes whose role in the transfer is not negligible. We now analyze the effect of each plasmonic mode in the population transfer by artificially truncating the basis choosing the number of modes in Fig. \ref{fig:fig7}.

Figure \ref{fig:fig7}(a) shows that an efficient transfer occurs for $\phi=0$ or $\phi=\pi$ (and around these angles) when the sole dipolar mode is taken into account. 
Involving more and more modes has a detrimental effect on the transfer for $\phi=\pi$ and reduces the angular width of transfer around $\phi=0$.

The blockade of transfer for $\phi=\pi$ can be thus explained by a destructive interference of the modes since the overlap between the modes $n$ of the two emitters is given by $\mu^{21}_n=(-1)^n$. On the other hand, for $\phi=0$,  $\mu^{21}_n=1$ results in a constructive interference.


\begin{widetext}
\begin{center}
\begin{figure}[!h]
\includegraphics[scale=0.9]{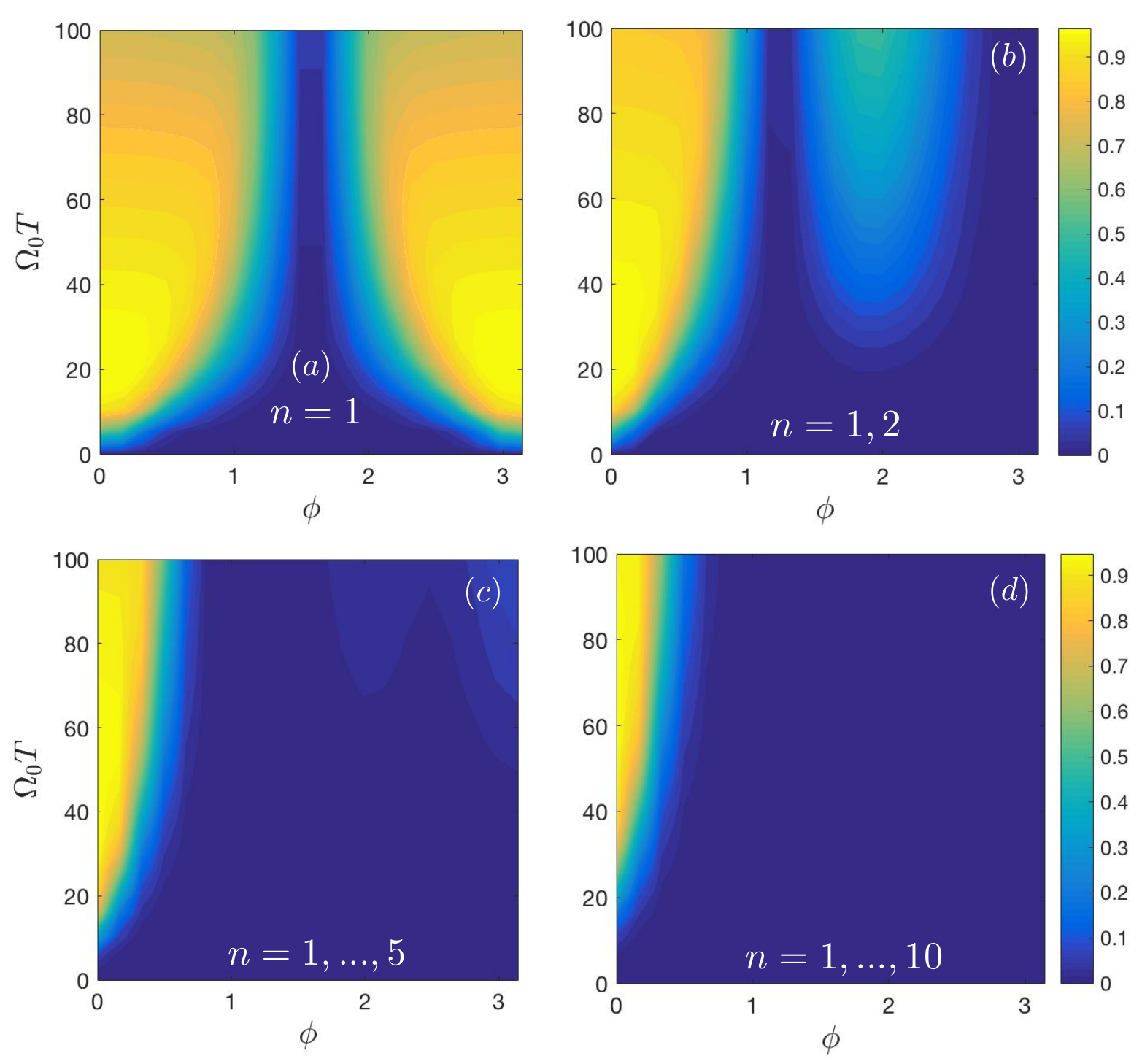}
\caption{
Countour plots of the transfer efficiency $|\langle g,f,0,0|\Psi(t=+\infty)\rangle |^2$ in function of the angular distance $\phi \in [0,\pi]$ and of the area of the laser pulses $\Omega_0T$, for the artificial truncation of the number of modes, as indicated in (a)-(d).}
\label{fig:fig7}
\end{figure}
\end{center}
\end{widetext}

\begin{figure}[!h]
\begin{center}
\includegraphics[scale=0.8]{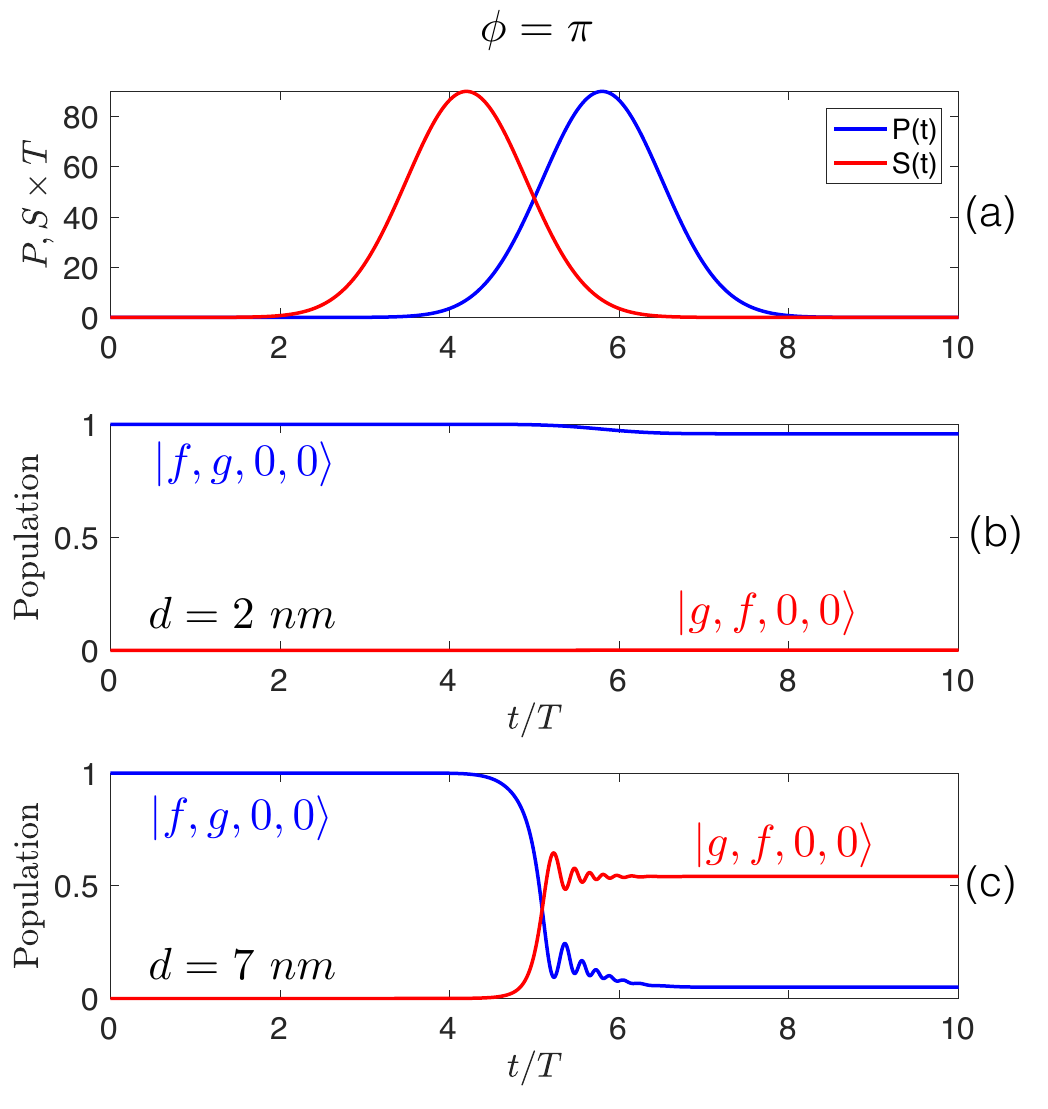}
\end{center}
\caption{Dynamics of the population transfer from $|f,g,0,0\rangle$ to  $|g,f,0,0\rangle$ for $\phi=\pi$ and $d=2$ nm, for which 25 modes are involved and  featuring the blockade (middle frame), and $d=7$ nm, for which 7 modes are involved (lower frame). Gaussian profile of the two delayed and counterintuitively ordered laser pulses (upper frame). The area of each pulse is $\Omega_0T=90$ and the delay is $\tau/T=0.8$, where $T\simeq 15$ ns.}
\label{fig:fig9}
\end{figure}


The distance between each emitter and the metallic surface determines the number of plasmonic modes effectively involved in the transfer.
The greater is the distance, the smaller is the number of the involved modes (see Fig. \ref{fig:fig1}). Increasing the distance from the metallic surface up to $7$ nm, for $\phi=\pi$, weakens the blockade: almost 50\% of the population is transferred to the target state in this case (see Fig. \ref{fig:fig9}).

\section{CONCLUSIONS}\label{IV}

We have constructed mode-selective effective models describing the interaction of $N$ QEs with the LSPs supported by a spherical metal nanoparticle in an arbitrary geometric arrangement of the QEs. We have developed a general formulation including the decomposition into orthogonal modes with the example of spherical symmetry.
We have shown that a Stimulated Raman Adiabatic Passage configuration allows population transfer, as an exchange of population between two QEs, when the QEs are located on the same side of the MNP and very closed to it. The transfer is blocked when the emitters are positioned at the opposite sides of the MNP by the destructive superposition of all the interacting plasmonic modes.
We have investigated the blockade effect in term of the numbers of modes involved in the process. For  QEs at the same side of the MNP, the coupling is magnified by all the modes. On the other hand, population transfer is blocked for QEs located at the opposite sides if many modes are involved as this is the case when the QEs are very closed to the MNP (a few nanometers).
When the dipolar modes are only involved, the blockade is prevented for $\phi=\pi$ and the transfer can occur if the coupling is sufficiently strong. For a spherical MNP, one cannot find a compromise between strong coupling and dipolar modes only involved that lead to efficient population transfer. One can thus anticipate that tubular geometry with an enhanced dipolar mode coupling would be more appropriate for such transfer with $\phi=\pi$.

A generalization to the case of $N$ emitters will offer the possibility to produce a N-qubit processor at the nanoscale, via closest neigbough effective coupling.

\begin{acknowledgments}
This work was supported by the French ``Investissements d'Avenir'' program, project ISITE-BFC / I-QUINS (contract ANR-15-IDEX-03).
We acknowledge additional support from the Labex ACTION program (ANR-11-LABX-01-01) and PLACORE (ANR-BS10-0007).
\end{acknowledgments}

\section*{Appendix A: Overlap Matrix}
We consider a set of $N$ vectors $ |c_i\rangle \in \mathcal{H}$, that span a subspace $\mathcal{H}_{\text{ind}} \subset \mathcal{H}$, whose dimension coincides with the number $N_{\text{ind}} \leq N$ of independent vectors in the defined set. We want to determine the dimension of this subspace. For this purpose, we define a one row array and its formal adjoint
\begin{equation}
\mathbf{C}:=\left[|c_1\rangle, ...,|c_N\rangle\right], \qquad \mathbf{C}^{\dagger}:=\begin{bmatrix}
\langle c_1| \\
\vdots \\
\langle c_N|
\end{bmatrix},
\label{Cset}
\end{equation}
and we define a rule of multiplication that is similar to the tensor product of two vectors:
\begin{equation}
\mathbf{C}^{\dagger}\mathbf{B}=\begin{pmatrix}
\langle c_1,b_1\rangle & \cdots & \langle c_1,b_N\rangle \\
\vdots & \ddots & \vdots \\
\langle c_N,b_1\rangle & \cdots & \langle c_N,b_N\rangle
\end{pmatrix}.
\end{equation}
We introduce the \textit{overlap matrix} $M$, called also \textit{metric matrix} or \textit{Gram matrix} corresponding to the set of vectors $|c_i\rangle$ as 
\begin{equation}
M:=\mathbf{C}^{\dagger}\mathbf{C}=\begin{pmatrix}
\langle c_1,c_1\rangle & \cdots & \langle c_1,c_N\rangle \\
\vdots & \ddots & \vdots \\
\langle c_N,c_1\rangle & \cdots & \langle c_N,c_N\rangle
\end{pmatrix}.
\label{overlapmatrixgen}
\end{equation}
It is an $N\times N$ Hermitian and positive semi-definite matrix as show below.

\vspace{0.2cm}
\noindent \textbf{Lemma A.1}: Let $\{|\varphi_i\rangle \}$ be an arbitrary set of $N$ vectors, then the eigenvalues of the overlap matrix $M$ with elements $M_{ij}:=\langle \varphi_i,\varphi_j\rangle$ are positive or zero. Furthermore, if the vectors $|\varphi_i\rangle$ are linearly independent, all the eigenvalues of $M$ are strictly positive and thus $M$ is invertible.

\vspace{0.1cm}
\textbf{Proof}: Let $\mathbf{v}^{k} \in \mathbb{C}^N$ be an eigenvector of $M$, i.e. $M\mathbf{v}^{k}=\lambda_k\mathbf{v}^{k}$ and define the following linear combination
\begin{equation}
|\phi^k\rangle :=\sum_i |\varphi_i\rangle v_i^k.
\end{equation}
Then
\begin{equation}
\begin{split}
\langle \phi^k,\phi^k\rangle &=\sum_{ij}\langle \varphi_i ,\varphi_j\rangle v^{k\ast}_iv^k_j=\sum_i v^{k\ast}_i\sum_j M_{ij}v^k_j \\&
=\sum_i v^{k\ast}_i\lambda_kv^k_j=\lambda_k\sum_i v^{k\ast}_iv^k_i=\lambda_k|\mathbf{v}^k|^2
\end{split}
\end{equation}
and thus
\begin{equation}
\lambda_k=\dfrac{||\phi^k||^2}{|\mathbf{v}^k|^2}\geq 0.
\end{equation}
The strict inequality follows from the fact that the linear independence of the $|\varphi_i\rangle$ implies that  $|\phi^k\rangle \neq 0$. $\square$

\vspace{0.2cm}

\noindent The \textit{rank} of a set of vectors $|c_i\rangle \in \mathcal{H}$, defined as the dimension of the subspace they span, is equal to 
\begin{itemize}
\item the maximal number of linearly independent vectors.
\item the dimension of the image of the linear map defined by 
\begin{eqnarray}
\mathbf{C}: \quad \mathbb{C}^N &\mapsto &\mathcal{H}\\
\mathbf{v} &\mapsto& \mathbf{C}\mathbf{v}=\sum_{i=1}^N |c_i\rangle v_i,
\end{eqnarray}
i.e. equal to $N$ minus the dimension of the kernel of this map.
\end{itemize}

\noindent \textbf{Lemma A.2}: A vector $\mathbf{v} \in \mathbb{C}^{N}$ satisfies the following condition
\begin{equation}
\mathbf{C}\mathbf{v}=|0\rangle \Longleftrightarrow M\mathbf{v}=\mathbf{0},
\end{equation}
where the vector $|0\rangle$ indicates the vector zero in $\mathcal{H}$ and $\mathbf{0}$ is the vector zero in $\mathbb{C}^{N}$. This means that the kernel of the map $\mathbf{C}$ is equal to the kernel of the matrix $M$.

\vspace{0.1cm}
\textbf{Proof}:

($\Rightarrow $): $M\mathbf{v}=\mathbf{C}^{\dagger}\mathbf{C}\mathbf{v}$ and thus $\mathbf{C}\mathbf{v}=\vert0\rangle$ implies $M\mathbf{v}=\mathbf{0}$.

($\Leftarrow $): If $M\mathbf{v}=\mathbf{C}^{\dagger}\mathbf{C}\mathbf{v}=\mathbf{0}$, it implies that
\begin{equation}\begin{split}
0&=\mathbf{v}\cdot \left(\mathbf{C}^{\dagger}\mathbf{C}\mathbf{v}\right)=\sum_{i,j=1}^N v^{\ast}_i\langle c_i,c_j\rangle v_j\\
&=\left\langle \sum_i v_i c_i,\sum_j v_j c_j\right\rangle = \langle \mathbf{C}\mathbf{v},\mathbf{C}\mathbf{v}\rangle
\end{split}
\end{equation}
and thus $\mathbf{C}\mathbf{v}=|0\rangle$. $\square$

\vspace{0.1cm}
\noindent \textbf{Lemma A.3}: The number $N_{\text{ind}}$ of linearly independent vectors $|\varphi_i\rangle$ is equal to the rank of the overlap matrix $M$, which is equal to the number of its non-zero eigenvalues.

\vspace{0.1cm}
\textbf{Proof}: The rank of \textbf{C} is equal to $N$ minus the dimension of the kernel of the map \textbf{C} and, from Lemma A.2, the dimension of this kernel is equal to the dimension of the eigenspace of $M$ corresponding to the eigenvalue $0$. 


\section*{Appendix B: L\"{o}wdin canonical orthonormalization for non linearly independent vectors}
L\"{o}wdin introduced two \textit{global} methods of orthonormalization of a vector: the canonical approach and the symmetric one. The adjective \textit{global} indicates that, given a set of non-orthogonal linearly independent vectors, both methods consider simultaneously all of them in the process of construction of the orthonormal set \cite{Lowdin}. We generalize the canonical method to the case in which the non-orthogonal vectors are not necessarily linearly independent. 

We introduce a set $\{|c_i\rangle\}$ of vectors and the corresponding overlap matrix $M$. The vectors $|c_i\rangle$ are not mutually orthogonal, i.e. $\langle c_i,c_j\rangle \neq 0$. We want to obtain from them an orthonormal set $\{|b_i\rangle\}$. The overlap matrix $M$ is Hermitian and so it can be diagonalized by a unitary matrix $T$
\begin{equation}
T^{\dagger}MT=D=\text{diag}(\lambda_1,...,\lambda_{N_{\text{ind}}},0,...,0),
\end{equation}
where the presence of the zero eigenvalues indicates the possibility that the vectors $|c_i\rangle$ are not linearly independent.
Defining the following matrix
\begin{equation}
D_{-1/2}:=\text{diag}(\lambda_1^{-1/2},...,\lambda_{N_{\text{ind}}}^{-1/2},0,...,0),
\end{equation}
according to the L\"{o}wdin canonical method, we can obtain an orthonormal set of vectors $\mathbf{B}$ 
\begin{equation}
\mathbf{B}:=\mathbf{C}TD_{-1/2},
\label{Blow}
\end{equation}
where, following the notation introduced in \eqref{Cset}, the form of $\mathbf{B}$ is
\begin{equation}
\mathbf{B}=\left[|b_1\rangle,...,|b_{N_{\text{ind}}}\rangle,|0\rangle,...,|0\rangle\right]
\end{equation}
and $\mathbf{C}$ is the one row array defined in \eqref{Cset}. 

The new set of vectors $\mathbf{B}$ satisfies indeed the orthonormality condition
\begin{equation}\begin{split}
\mathbf{B}^{\dagger}\mathbf{B}&= D_{-1/2}T^{\dagger}MTD_{-1/2}\\
&=D_{-1/2}DD_{-1/2}\\
&=\text{diag}(\openone_{N_{\text{ind}}},0,...0),
\end{split}
\end{equation}
i.e. $\langle b_j, b_j^{'}\rangle = \delta_{jj^{'}}$, for all $j,j^{'} \in [1,N_{\text{ind}}]$.

According to \eqref{Blow}, the vectors $|b_j\rangle\in{\cal H}$ can be written in terms of the original ones in the following explicit form
\begin{equation}
|b_j\rangle :=\begin{cases}
\lambda_j^{-1/2}\sum_{i=1}^N|c_i\rangle T_{i,j} \qquad&\text{for } j\in [1,N_{\text{ind}}],\\
|0\rangle  \qquad&\text{for } j \in [N_{\text{ind}}+1,N]
\end{cases}
\label{blowsing}
\end{equation}


\noindent The relations \eqref{Blow} and \eqref{blowsing} can be inverted to express the original vectors in terms of the orthonormal basis:
\begin{equation}
\mathbf{C}:=\mathbf{B}D^{1/2}T^{\dagger},
\label{sing_v_dec}
\end{equation}
i.e.
\begin{equation}
|c_i\rangle :=\sum_{j=1}^{N_{\text{ind}}}\lambda^{1/2}_j T_{i,j}^{\ast}|b_j\rangle  \ \ \ \ \ \ i \in [1,N].
\label{cl}
\end{equation}
We remark that \eqref{sing_v_dec} has the form of a  {\it Singular Value Decomposition} 
extended for vectors $| c_{i}\rangle$ in an infinite dimensional Hilbert space \cite{Wilkinson-Algebraic-Eigenvalue-Problem-book}\cite{Trefethen-Bau-Numerical-Linear-Algebra}\cite{Numerical-Recipes}. The canonical L\"owdin orthonormalization can thus be viewed as the inverse of the Singular Value Decomposition of $\mathbf{C}$.

\noindent To prove \eqref{cl}, we start from the definition \eqref{blowsing}, obtaining
\begin{equation}\begin{split}
\lambda_{j}^{1/2}|b_j\rangle &:=\begin{cases}
\sum_{i=1}^N|c_i\rangle T_{i,j} \qquad & \text{for } j\in [1,N_{\text{ind}}],\\
|0\rangle \qquad & \text{for }  j \in [N_{\text{ind}}+1,N]
\end{cases}\\
&=\sum_{i=1}^N|c_i\rangle T_{i,j} \qquad \text{for } j \in [1,N],
\end{split}
\label{boh}
\end{equation}
where the second equality follows from Lemma A.2: for $j \in [N_{\text{ind}}+1,N]$, the columns of the overlap matrix are the eigenstates $\mathbf{t}_j$ corresponding to the eigenvalue $0$, i.e. $M\mathbf{t}_j=0$, and, by Lemma A.2, $\mathbf{C}\mathbf{t}_j=|0\rangle$.

\noindent Multiplying both sides of \eqref{boh} by $T^{\ast}_{i^{'},j}$ and summing over $j$ we obtain
\begin{equation}\begin{split}
\sum_{j=1}^N|b_j\rangle\lambda_j^{1/2}T^{\ast}_{i^{'}j}&=\sum_{i=1}^N|c_i\rangle \sum_{j=1}^{N} T_{i,j}T^{\ast}_{i^{'},j}\\&=\sum_{i=1}^N|c_i\rangle \delta_{i,i^{'}}=|c_{i'}\rangle,
\end{split}
\end{equation}
which is the singular value decomposition \eqref{sing_v_dec}.
\nocite{*}

%

\end{document}